\documentclass[11pt,letterpaper]{article}
\pdfoutput=1
\usepackage{enumerate}
\usepackage{mathtools}
\usepackage{axodraw2}
\usepackage{amscd}
\usepackage{amsmath}
\usepackage{slashed}
\usepackage{color}
\usepackage{gensymb}
\usepackage{tikz}
\usetikzlibrary{shapes,arrows,positioning,automata,backgrounds,calc,er,patterns}
\usepackage{amscd}
\usepackage{multirow}
\usepackage{float}
\usepackage{jheppub} 
\usepackage{cancel}
\usepackage{fontawesome}

\def\qmax{10}

\newcommand{\minus}{\scalebox{0.5}[1.0]{$-$}}

\title{A $\nu$ Supersymmetric Anomaly-free Atlas}

\author[a]{B.C. Allanach,}
\author[a]{Maeve Madigan,\footnote{Corresponding author.}}
\affiliation[a]{DAMTP, University of Cambridge, Wilberforce Road, Cambridge, 
CB3 0WA, United Kingdom}
\author[b]{Joseph Tooby-Smith}
\affiliation[b]{Cavendish Laboratory, University of Cambridge, J J Thomson
  Ave, Cambridge, CB3 0HE, United Kingdom} 

\emailAdd{B.C.Allanach@damtp.cam.ac.uk}
\emailAdd{mum20@cam.ac.uk}
\emailAdd{jss85@cam.ac.uk}

\abstract{Extensions of the minimal supersymmetric standard model (MSSM) gauge group
abound in the literature. Several of these include an additional $U(1)_X$ gauge group.
Chiral fermions' charge assignments under $U(1)_X$
are constrained to cancel local anomalies in the extension
and they determine the
structure and phenomenology of it. 
We provide all anomaly-free charge assignments up to
a maximum absolute charge 
of $Q_\text{max}=\qmax$, assuming that the chiral superfield content of
the model is that of the MSSM plus up to
three Standard Model (SM) singlet superfields.
The fermionic components of these SM singlets may play the r\^{o}le of
right-handed neutrinos, whereas one of the scalar components may play the
r\^{o}le of the flavon, spontaneously breaking $U(1)_X$. 
Easily scanned lists of the charge assignments are made publicly
available on {\tt Zenodo}.
For the case where no restriction is placed upon $Q_\text{max}$, 
we also provide an analytic parameterisation of the general
solution using simple techniques from algebraic geometry.}

\begin{document}
\maketitle
\flushbottom

\section{Introduction}
\label{sec:intro}

Quantum field theories of vector bosons are notoriously problematic unless
they arise from gauge symmetries, whence non-renormalisability and
non-unitarity can be tamed. It is thus imperative that
the gauge symmetry of the renormalisable ultra-violet completion of any such
model should not contain any quantum field theoretic gauge \emph{anomalies}, where
quantum corrections spoil the gauge symmetry that was imposed upon the tree-level theory.
The Standard Model (SM) itself is anomaly-free and can thus remain a self-consistent theory
up to very large renormalisation scales. Despite this, there are good reasons to expect the SM
to be an effective field theory resulting from decoupling other fields.
Many reasons have been invoked to motivate extending the Lie
algebra\footnote{We shall 
  refer to the Lie algebra (as opposed to the Lie group) in
  $\mathfrak{mathfrak}$ script.}
$\mathfrak{sm}:=\mathfrak{su}(3)\oplus\mathfrak{su}(2)\oplus\mathfrak{u}(1)_Y$
of the Standard Model 
(SM) by a spontaneously broken gauged $\mathfrak{u}(1)_X$ summand, for
example. Such 
extensions have been used to explain measurements of the anomalous magnetic
moment of the 
muon~\cite{Heeck:2011wj}, to provide axions~\cite{Berenstein:2010ta} or leptogenesis~\cite{Chen:2011sb}, to provide
fermion masses through the 
Froggatt-Neilsen mechanism~\cite{Froggatt:1978nt}, or explain measurements of the
$b\rightarrow s l^+ l^-$ transition which are currently in tension with SM predictions~\cite{Altmannshofer:2014cfa,Alonso:2017uky,Bonilla:2017lsq,Bhatia:2017tgo,Ellis:2017nrp,Allanach:2018lvl,Allanach:2019iiy,Greljo:2021xmg,Davighi:2021oel}. In general, the $X$ charge
assignments of the models can be family dependent, resulting in
family-dependent couplings of a resulting massive $Z^\prime$ vector boson.
Indeed, in several applications (the last two in our aforementioned list) it
is a necessary requirement that the $X$ charges are family dependent, since
the symmetry and the $Z^\prime$ are respectively used to explain family non-universal effects. 

In $\mathfrak{u}(1)_X$ extensions, the phenomenology of the $Z^\prime$ is
often key and is dictated by
the integer $X$ charges of the other fields in the model (integer $X$ charges
results from an implicit assumption that the extension is compact).
The $X$ charges of the chiral fermions in particular dictate the
contribution to perturbative local anomalies of such models. 
There is therefore a non-trivial cross-over between the extensions'
phenomenology and anomaly cancellation via the chiral fermions' charge
assignments. Unfortunately, in general, with a fixed chiral fermion content,
anomaly cancellation conditions (ACCs) are difficult to solve, the number theory state-of-the art
being the solution of a single cubic in three unknown integer parameters~\cite{Mordell}. 

Some recent progress has been made in this direction, however. In Ref.~\cite{Costa:2019zzy}, the
gravitational and gauge anomalies of a pure $U(1)$ gauge symmetry
(i.e.\ with no SM gauge group but with charged chiral fermionic fields) were
solved analytically for the charges 
of \emph{a priori} fixed numbers of chiral fermions via an ingenious
algebraic method\footnote{{The algebraic approach was {partially} extended to $U(1)^n$ gauge symmetries in Ref.~\cite{Costa:2020dph}.}}; this was soon understood from a geometric
perspective~\cite{Allanach:2019gwp} by using a theorem due to Mordell~\cite{Mordell}. 
Similar geometric methods were employed to find an analytic solution to the
more difficult problem of $\mathfrak{sm}\oplus\mathfrak{u}(1)$ anomaly-free charge assignments in
the specific case of SM fermion content, plus three right-handed (RH) neutrinos
(i.e.\ SM-singlet chiral fermion fields which may carry $X$ charge)~\cite{Allanach:2020zna}.
The number of solutions is formally infinite,\footnote{One way of seeing this
  is to set the $X$ charges of the first family of particles to
  be equal to their hypercharges, the second family to be equal to some
  integer multiplied by baryon
  number minus lepton 
  number $B-L$, and the third family to have zero
  charge. Any such charge assignment solves the anomaly cancellation
  conditions. Since there are an infinite number of constants we can multiply
  the second family by, each of which leads to a distinct chiral solution, there
  are an infinite number of solutions.} 
unlike the case of semi-simple
SM extensions with identical fermionic field content, {where there is a list of
340}~\cite{Allanach:2021bfe}.
Unfortunately, the geometric methods employed only solve a small family of
similar cases and cannot be deployed on general chiral fermionic contents.
Furthermore, the analytic solution, whilst of intrinsic interest in and of
itself, comes with a significant drawback for model-builders interested in
using it: each charge is parameterised in terms of a fourth-order polynomial of
integer parameters. Whilst it is easy to input these parameters and
achieve 
anomaly-free charges, model builders often want to 
fix a function of them to certain values
for
phenomenological purposes, but this is a difficult and currently
unsolved problem, because it
involves solving a system of coupled fourth-order diophantine equations. 

Fortunately, when appropriately employed, computers
come to the rescue of the reverse-engineering model builder.
In an $\mathfrak{sm}\oplus\mathfrak{u}(1)$ `anomaly-free
atlas'~\cite{Allanach:2018vjg},
all solutions of the ACCs for integer charges between 
-10 and 10 for
18 chiral fermion gauge representations in the SM plus three RH
neutrinos were found by a scan.\footnote{This strategy has also recently been used for the case of
  $U(1)$ gauge theory with different numbers of Weyl fermions, in a search
  for scotogenic models~\cite{Wong:2020obo}.}
Cases which are in a sense equivalent (where the charges differ by a common
multiple which can be absorbed into the $\mathfrak{u}(1)_X$ gauge coupling, or which
differ by a permutation of the family indices within a \emph{species} - fields
which have identical SM 
representations) were only counted once (and aside from some rare cases, only
scanned over once). 
Anomaly-free solutions
are scarce: only roughly one in $10^{9}$ 
was anomaly-free from the whole
sample. 
The list of anomaly-free fermionic
charge assignments was made publicly available. It is a list of over
$21\ 000\ 000$ solutions that is easy and quick to search through and filter with
the aid of a simple computer program. As such, it is user friendly for
would-be $U(1)_X$ gauge extension model builders who can search through the
list and filter for charge assignments with various desired
properties. The charges are 
limited in \emph{height} (the maximum absolute value of a charge in any
solution), but have the advantage of being easily useable provided one can 
adapt or write a simple computer program that reads the list in and filters it. 

Heretofore, there has been no similar list made for supersymmetric (SUSY) models.
SUSY model building has several motivations, the
primary one being that it does not suffer from the technical hierarchy
problem, where radiative corrections to the Higgs mass tend to drag it up to
the largest fundamental energy scale (for example the Planck mass $\sim
10^{19}$ GeV) divided by a loop factor. There are other motivations for
supersymmetry too, for
example, in an $\mathcal{N}=1$
supersymmetrisation of the SM (the MSSM), the
experimental measurements of the gauge couplings agree with 
the gauge coupling unification condition predicted by SUSY grand
unified theories. When one includes an extra multiplicative discrete symmetry such as 
$R-$parity or matter parity\footnote{Matter parity is defined as
  $(-1)^{3(B-L)}$, where $B$ is baryon number and 
$L$ is lepton number, whereas $R-$parity is defined as $(-1)^{3(B-L)+2s}$, where $s$ is
  spin.}
the MSSM possesses a stable particle which,
depending upon parameters,  
has the correct properties to constitute the universe's dark matter and
potentially dangerous proton decay processes are suppressed.
Particular examples of $\mathfrak{u}(1)_X$ gauge extensions of the MSSM
can combine the aforementioned phenomenological benefits of a $Z^\prime$ with
those of SUSY models. Some of these
have appeared in the literature, for example see
Refs.~\cite{Demir:2005ti,Barger:2008wn,Duan:2018akc,sym13020191,Ashmore:2021xdm,Frank:2021nkq}.

It is our intention here to extend the original 
non-SUSY anomaly-free atlas to the SUSY case and make a
new list (a `$\nu$ SUSY anomaly-free atlas') available to interested
SUSY $\mathfrak{u}(1)_X$-extension model 
builders and others. We shall include the addition of up to three MSSM-singlet chiral
superfields: the fermionic components of all or some of these can play the
r\^{o}le of RH neutrinos, 
resulting in tiny neutrino masses via the see-saw mechanism (below, we call
this model the $\nu$MSSM).
The scalar component of one of these MSSM-singlet chiral superfields is
expected to play the r\^{o}le of the flavon, which has a necessarily non-zero
$X$ charge and acquires a vacuum 
expectation value, spontaneously breaking $U(1)_X$.
One might expect that one of the SM-singlet fields must therefore have a non-zero
$\mathfrak{u}(1)_X$ charge, unlike the non-SUSY case, where the charges of the
flavon and all SM-singlet fermions were \emph{a priori} unconstrained.
However, we won't impose this condition because the field content of the model
can easily be extended in a way that does not change the ACCs but which
effectively removes the condition, as we shall explain below.
A functional difference to the original non-SUSY anomaly-free atlas is
the appearance of the Higgsino partners of the two MSSM Higgs doublets,
augmenting the number of Weyl fermion $SU(2)$ gauge representations by two. This
therefore extends the original list of 18 $X$ charges to 20.
In case a height larger than 10 is required, we will also provide a general analytic solution to
the anomaly cancellation conditions. This relies on using the same geometric
framing in which the SM-plus-3 RH neutrino case was
solved~\cite{Allanach:2020zna}; we 
take the opportunity to demonstrate a new technique to solve such problems,
although the technique used in Ref.~\cite{Allanach:2020zna} would also have
worked.  

The paper proceeds as follows: in \S\ref{sec:ACC}, we describe the anomaly
cancellation conditions relevant for the Lie algebra
$\mathfrak{mssm}\oplus\mathfrak{u}(1)_X$,
and a chiral superfield content of the $\nu$MSSM\@. In
\S\ref{sec:comp}, we describe the computational scan and how the solutions are listed
and ordered, giving the number of solutions found up to a height of 10. 
We provide an analytic method of solution in \S\ref{sec:analytic},
along with a parameterisation of the solution. Various consistency checks of the
solutions are described in \S\ref{sec:checks}: some are checks solely of the
numerical solutions, some are of the analytic solution and some are checks of
the analytic solution versus the numeric solutions.
Some initial filters of the numerical solutions (chosen for specific
phenomenological reasons) are explored in
\S\ref{sec:pheno}. We provide a summary of the paper and a discussion in \S\ref{sec:summary}.
 
We list chiral fermionic fields
in
the representations displayed in Table~\ref{tab:convs}. 
As previously mentioned, the left-handed fermionic fields contained within the two Higgs chiral
superfields provide a new feature as regards the ACCs. We note here that the
fermionic components of the
chiral superfields $H_d$ and $L_i$ have identical representations under the
SM gauge Lie algebra, but the fermionic component of $H_d$ may
or may not be discriminated by a
different quantum number under an imposed symmetry such as matter parity or
$R-$parity. 

We have thus augmented the MSSM, as far as the fermionic $X$ charges go,
by 20 parameters which we write in a 20-tuple
\begin{align}
{\bf X}:=\{&X_{Q_1},X_{Q_2},X_{Q_3},X_{n_1},X_{n_2},X_{n_3},X_{e_1},X_{e_2},X_{e_3},X_{u_1},X_{u_2},X_{u_3},X_{d_1},
\nonumber \\ &X_{d_2},X_{d_3},X_{L_1},X_{L_2},X_{L_3},X_{H_d},X_{H_u}\}. \label{tuple}
\end{align}
We take it as understood that, for the case where $R-$parity is \emph{not} a
symmetry of the theory, we modify (\ref{tuple}) such that $X_{H_d}$ is
merged with $X_{L_i}$ to form $X_{L_\alpha}$, where $\alpha \in \{1,2,3,4\}$.
For now though, we shall continue the discussion where $H_d$ is discriminated
from $L_i$ by a discrete symmetry. Since the gauge extension is here assumed to
be compact, ${\bf X}$ is \emph{a priori} valued in $\mathbb{Z}^{20}$.

\begin{table}[h!]
\begin{center}
  \begin{tabular}{|r|cccc|}\hline
    \multicolumn{5}{|c|}{Fermions} \\ \hline 
     & $\mathfrak{su}(3)$ & $\mathfrak{su}(2)_L$ & $\mathfrak{u}(1)_Y$ & $\mathfrak{u}(1)_X$ \\
    \hline
    LH quark doublets $Q_i$ & 3 & 2 & 1 & $X_{Q_i}$ \\
    RH neutrinos $n_i$ & 1 & 1 & 0 & $X_{n_i}$ \\
    RH charged leptons $e_i$ & 1 & 1 & -6 & $X_{e_i}$ \\
    RH up quarks $u_i$ & 3 & 1 & 4 & $X_{u_i}$ \\
    RH down quarks $d_i$ & 3 & 1 & -2 & $X_{d_i}$ \\
    LH lepton doublets $L_i$ & 1 & 2 & -3 & $X_{L_i}$ \\
    LH down-type Higgsino $\tilde H_d$ & 1 & 2 & -3 & $X_{H_d}$ \\
    LH up-type Higgsino $\tilde H_d$ & 1 & 2 & 3 & $X_{H_u}$ \\        
    \hline
    \multicolumn{5}{|c|}{Chiral superfields} \\ \hline 
    $\hat Q_i$ & 3 & 2 & 1 & $X_{Q_i}$ \\
    $\hat N_i^c$ & 1 & 1 & 0 & $-X_{n_i}$ \\
    $\hat E_i^c$ & 1 & 1 & 6 & $-X_{e_i}$ \\
    $\hat U_i^c$ & $\bar 3$ & 1 & -4 & $-X_{u_i}$ \\
    $\hat D_i^c$ & $\bar 3$ & 1 & 2 & $-X_{d_i}$ \\
    $\hat L_i$  & 1 & 2 & -3 & $X_{L_i}$ \\
    $\hat H_d$ & 1 & 2 & -3 & $X_{H_d}$ \\
    $\hat H_d$ & 1 & 2 & 3 & $X_{H_u}$ \\            
    \hline
  \end{tabular}
    \end{center}
  \caption{\label{tab:convs} Conventions for field content with
    representations under the 
    gauge Lie algebra. RH stands for right-handed and LH stands for
    left-handed. $i\in \{1,2,3\}$ is a family index. Note that we have re-scaled a more conventional hypercharge
  assignment by a factor of 6 to make all hypercharges setwise coprime
  integers. Such a re-scaling can be
  absorbed into the hypercharge gauge coupling. ${}^c$ denotes
  charge
  conjugation on the scalar and fermionic components of the chiral superfield.}
  \end{table}

\section{$\mathfrak{u}(1)_{X}$ Extension of the MSSM Lie Algebra}
\label{sec:ACC}
\subsection{Anomaly cancellation conditions} \label{sec:ACCconditions}
 
The MSSM \emph{per se} is anomaly free. With the addition of {$\mathfrak{u}(1)_X$}, local
 anomalies persist unless {\bf X} satisfies the 
ACCs\footnote{Note that where necessary, we discriminate between the gauge Lie
  algebra, which is equivalent to $\mathfrak{sm}\oplus {\mathfrak u}(1)_X$ and
  the MSSM$\times U(1)_X$ gauge group, which 
  is strictly only determined up to certain quotients, but this
  does not affect any of our discussion.}
\begin{align}
\mathfrak{su}(3)^{2} \oplus \mathfrak{u}(1)_{X}:\  &\sum_{i=1}^{3} (2 X_{Q_{i}}
- X_{u_{i}} - X_{d_{i}}) = 0, \label{lin1}\\ \mathfrak{su}(2)^{2} \oplus
\mathfrak{u}(1)_{X}:\  &\sum_{i=1}^{3}(3 X_{Q_{i}} +X_{L_{i}})+ X_{H_{d}} +
X_{H_{u}} = 0,
\end{align}
\begin{align}
\mathfrak{u}(1)_X\textrm{-gravity} :\  &\sum_{i=1}^{3}(6
X_{Q_{i}}- X_{n_{i}} - X_{e_{i}} - 3 X_{u_{i}} - 3 X_{d_{i}} +2X_{L_i})
+2X_{H_d} +2 X_{H_{u}} = 0,\label{lin2}\\ \mathfrak{u}(1)_{X}^{3} :\ 
&\sum_{i=1}^{3} (6 X_{Q_{i}}^{3}- X_{n_{i}}^{3} - X_{e_{i}}^{3} - 3
X_{u_{i}}^{3} - 3 X_{d_{i}}^{3}+2X_{L_i}^3)+2 X_{H_d}^3+ 2 X_{H_{u}}^{3} =
0, \label{cubic}\\ \mathfrak{u}(1)_{X}^{2} \oplus \mathfrak{u}(1)_{Y} :\ 
&\sum_{i=1}^{3} (X_{Q_{i}}^{2} - 2 X_{u_{i}}^{2} + X_{d_{i}}^{2} +
X_{e_{i}}^{2}+X_{L_{i}}^{2} ) + X_{H_{d}}^{2}- X_{H_{u}}^{2} =
0, \label{quadratic}\\ \mathfrak{u}(1)_{Y}^{2} \oplus \mathfrak{u}(1)_{X}:\ 
&\sum_{i=1}^{3} (X_{Q_{i}} -6 X_{e_{i}}- 8 X_{u_{i}} - 2X_{d_{i}}+3X_{L_i}) +
3X_{H_d}+ 3X_{H_{u}} = 0.
\label{eq:ACCs}
\end{align}
These ACCs inherit some in-practice physical equivalences between
${\mathfrak{u}}(1)_X$ extensions related by the following operations:
\begin{enumerate}[(i)]
  \item Permutation of family indices within each species, since this is
    really just a change of basis.  \label{perm}
  \item ${\bf X}\rightarrow a {\bf X}$, where $a\in \mathbb{Q}\backslash\{0\}$,
    when
    the gauge coupling only appears in the Lagrangian multiplied by a $U(1)_X$
    charge, since
    the
    $U(1)_X$ gauge coupling may be simultaneously re-scaled by $1/a$ 
    resulting in no substantive change. This is displayed by
    the fact that the ACCs are homogeneous. \label{rescale}
  \item ${\bf X}\rightarrow {\bf X}+ y{\bf Y}$, where ${\bf Y}$ is the
    20-tuple of fermionic field hypercharges (in the same field ordering as
    ${\bf X}$) and $y\in \mathbb{Z}$. Resulting from a group outer
    automorphism, this change in fermionic representations can be accounted
    for by a redefinition of gauge fields~\cite{Costa:2020dph}. \label{hyper}
  \end{enumerate}
Ideally, we wish to record exactly one entry in a list for each physically
inequivalent charge assignment.  Together, (\ref{rescale}) with (\ref{hyper})
imply that
we should regard ${\bf X} \rightarrow x {\bf X} + y {\bf Y}$ as an
equivalent theory, where $x\in \mathbb{Q}\backslash\{0\}$ and $y\in \mathbb{Q}$.
Unfortunately, we have not found an easy enough and fast enough method of
incorporating this, implying
that there will
remain a few physically 
equivalent charge assignments in any anomaly-free list that we produce. 
The necessary existence of these will end up providing us with a check of our
computer program in \S\ref{sec:checks}.
In any case, such equivalent charge assignments are rare, and we do
not foresee particular problems resulting from their presence in our list.
From now on, we refer to
`inequivalent' solutions to implicitly mean inequivalent under conditions (\ref{perm})
and (\ref{rescale}) only. 

To incorporate (\ref{perm}), we take
the convention that the family indices in ${\bf X}$ are such
that, for each species $S \in \{ Q, n, e, u, d, L \}$, $X_{S_1} \leq
X_{S_2} \leq X_{S_3}$ (for the case without additional discrete symmetries to
distinguish $H_d$ and $S=L$, $X_{S_3} \leq X_{S_4}$ as well).
To take (\ref{rescale}) into account, all integers in the tuple must be setwise coprime but
note that this still does not implement the equivalence with 
${\bf X}^\prime:=\{-X_{Q_3},-X_{Q_2},-X_{Q_1}, \ldots,-X_{H_d},-X_{H_u}\}$. In
order to
only list one instance of ${\bf X}, {\bf X}^\prime$, we must define a
condition that unambiguously picks one of them: here, we use the \emph{lexicographically 
  smaller}\/ tuple.\footnote{Lexicographical ordering is a much simpler condition
than the one used in the original anomaly-free
atlas~\cite{Allanach:2018vjg}.}
An $n-$tuple $a=\{a_1, \ldots, a_n\}$ is
lexicographically smaller than another $n-$tuple $b=\{b_1, \ldots, b_n \}$
(written as $a<b$) if and only if an
$i\in \{1,\cdots, n\}$ exists such that $a_i<b_i$ and $a_j=b_j$ for all $j\in \{1,\cdots,i-1\}$.

\subsection{Symmetry breaking}

Since we are not empirically aware of a long-range force that can be
attributed to an unbroken $U(1)_X$ gauge symmetry, we suppose that it must be
spontaneously broken. We further assume that it is broken by (at least) one of
the scalars ${\tilde \nu}_{R_i}$ contained in the
SM-singlet chiral superfields $N_i^c$,
so that it does not break the SM gauge
symmetry.
In order for a ${\tilde \nu}_{R_i}$ field to play this r\^{o}le, by Goldstone's
theorem
it must possess a non-zero $X$ charge. Typically, such a field is
called a \emph{flavon}. Let us denote it for the purposes of the current
discussion, as $\theta$. 
Contrary to the non-SUSY case,
we obtain a contribution
to
the ACCs through its fermionic superpartner $\tilde{\theta}$, the flavino.
However, we will still solve
the ACCs as given above assuming three SM-singlet chiral superfields only: $N_{i}^c$, where
$i\in \{1,2,3\}$. The reasons for not explicitly adding to this number (for example by adding
one more SM singlet chiral superfield) are twofold: firstly, we find 
practical barriers with four (or more) SM-singlets; the $\nu$ SUSY
anomaly-free atlas would take too long to compute and would take up too much
disk space to store for the desired height of 10. Secondly, by sticking to
three SM-singlet chiral superfields, we are able to find an analytic solution
to the ACCs.

In principle, the requirement that
at least one $X_{n_{i}} \neq 0$
would allow us to reduce the domain of $X$ charges
considered in our computational search below, although not by much.
We choose not to restrict the domain of $X$ charges in this way however, since  
one could augment our model by
\textit{two} additional scalar singlets $\theta_{1}$ and $\theta_{2}$ such
that $X_{\theta_{1}} + X_{\theta_{2}} = 0$.  The
contributions from $X_{\theta_{1}}$ and $X_{\theta_{2}}$ would cancel in the
ACCs, leaving the ACCs above unmodified.  This type of extension is
commonly used, for example, in $U(1)_{B-L}$ extensions of the MSSM
\cite{Barger:2008wn}. More generally one can add several SM-singlet chiral 
superfields which satisfy the pure
${\mathfrak{u}}(1)$ anomaly equations and thus cancel out of the
ACCs~\cite{Costa:2019zzy,Allanach:2019gwp}. We also note that to set a superfield's
charge to zero has the same effect \emph{on the ACCs} as would removing the
superfield (or at least its fermionic component) entirely from the model.

With the constraints (or lack thereof) listed above,
our inequivalent numerical set of solutions will have
the following subsets:
\begin{itemize}
\item The SM plus three RH neutrinos corresponds to the subset
  with $X_{H_u}=X_{H_d}=0$ (since this is equivalent to the model obtained by
  removing the $H_u$ superfield and the $H_d$ Higgsino from our current set-up). 
\item The MSSM with up to three $U(1)_{X}$-charged RH neutrino chiral
  superfields, where the $U(1)_{X}$ is broken by (at least) one of the RH sneutrinos,
  is the subset where at least one $X_{n_i}\neq 0$.
\item The MSSM with three RH neutrinos and two additional scalars $\theta_{1}$ and
  $\theta_{2}$ charged such that $X_{\theta_{1}} + X_{\theta_{2}} = 0$.
  This is a possibility for the subset with $X_{n_1}=X_{n_2}=X_{n_3}=0$.
  In
  particular, this option includes $U(1)_{X} = U(1)_{B-L}$ extensions of the
  MSSM. 
\end{itemize}

\section{Numerical Solutions up to a Height of 10\label{sec:comp}}
We produce a list of solutions to the ACCs using a  modification of the
computer program that was used to produce the original anomaly-free
atlas~\cite{Allanach:2018vjg}. To search for 
solutions, we scan over integer  values in the domain
$|X_i| \leq Q_\text{max}$.  Each set of solutions can then be classified by
$Q_\text{max}$, its maximum possible height. Note that in this definition, a list of solutions
$Q_\text{max}=N$ also contains all solutions with $Q_\text{max} < N$.

There are \emph{a priori}\/
$(2 Q_\text{max} + 1)^{20}$
solutions to be checked as solutions to the ACCs.
Time is the biggest limiting factor in our search for solutions.  The
original computer program is
described in detail in \S3 of Ref.~\cite{Allanach:2018vjg}, where we
direct the curious reader. Here, we find it expedient to not discriminate \emph{a priori} between $H_d$ and $L_i$: within the program, we therefore remove
the explicit $H_d$ charge, replacing it by $L_4$, equivalent to considering
the theory without a discrete symmetry to distinguish them.
The search for solutions is sped up by removing some equivalent solutions from
the scan, and by using the four linear ACCs to directly fix the values of four
of the
charges.  This still leaves us with a large solution space to consider,
as compared to the original non-supersymmetric anomaly-free atlas.  We further
improve the speed by parallelising 
the three outer loops (i.e.\ over $X_{Q_{1}}, X_{Q_{2}}$ and $X_{Q_3}$).  For $Q_\text{max} \leq 4$
the parallelisation improvement is minimal, but for $Q_\text{max}
\geq 5$ we find that this step is necessary to produce solutions in a reasonable
amount of time.

\subsection{Binary search algorithm \label{sec:bin}}
The output of the computer program for 
$Q_{\mathrm{max}}=10$ is a large lexicographically ordered list (the ASCII file
is around $125$ Gb in 
size) of inequivalent solutions which solve the ACCs, each one
comprised of a line made of the 20 integers which form ${\bf X}$.
We have not assumed a discrete symmetry that distinguishes $L_i$ from $H_d$
in the output and so we have four
$X_{L_\alpha}$ charges listed. 
This file
forms one of the two most important outputs of the present paper (the other
being the analytic solution for any height described in \S\ref{sec:analytic}). 
As mentioned in \S\ref{sec:intro}, we envisage that our output file may be used by
supersymmetric model builders by scanning through it with a computer program
and filtering the results. Since the file is so large though,
we have facilitated the 
decrease of the complexity of algorithms used to analyse the file in order to
speed them up.
The fact that
our list of solutions is ordered lexicographically means that one can take
advantage of 
the binary search algorithm. This reduces the complexity of finding a solution
in the list from $\mathcal{O}(n)$ to $\mathcal{O}(\log n)$. One usually has an
intuitive understanding of the binary search algorithm since it is  roughly
how one usually finds numbers in a phone book or words in a
dictionary, as follows. 
Let us say we have a solution we want to find in our list
or show that it does not exist in the list. The binary search algorithm goes
half-way down the list and determines if our solution is less then or equal
to the solution at the half-way point. If it is present in the first half then we throw away the
second 
half and keep the first, and if it isn't we discard the first half and keep
the second. This is then repeated until a list with a single item which will
(if it exists) match the one we are trying to find. 

\subsection{Output \label{sec:out}}
\begin{table}
\begin{center}
  \begin{tabular}{|c r r r r|} 
 \hline
 $Q_\text{max}$ &  \# $\nu$MSSM &  \# $R_p\nu$MSSM & \# Non-SUSY  & \# No ACC condition \\ [0.5ex] 
 \hline\hline
 1 & 111 &267 & 37 & $4.5\times 10^6$\\ 
 \hline
 2 &  2\,321&  6\,882 &357 & $2.3\times 10^{10}$ \\
 \hline
 3 & 44\,212 & 143\,707 &4\,115 & $8.6\times 10^{12}$ \\
 \hline
 4 & 401\,129 & 1367\,991 &24\,551 & $8.2 \times 10^{14} $\\
 \hline
  5 &  2\,582\,166 &9\,063\,191 & 111\,151 & $3.3 \times 10^{16}$ \\
   \hline
  6 &  13\,553\,325  &48\,681\,027  & 435\,304 & $7.6\times 10^{17}$ \\
 \hline
  7 &  54\,699\,483  & 199\,275\,965&  1\,358\,387 & $1.1 \times 10^{19}$ \\
 \hline
  8 &  185\,454\,955  & 682\,827\,818 & 3\,612\,733 & $1.2 \times 10^{20}$ \\
 \hline
  9 & 598\,267\,488  & 2\,224\,178\,673& 9\,587\,084 & $1.0\times 10^{21}$\\ 
  \hline
  10 & 1\,628\,002\,737  & 6\,094\,894\,134 & 21\,546\,919 & $6.8 \times 10^{21}$\\
 \hline
\end{tabular}
\end{center}
  \caption{Number of inequivalent anomaly-free charge assignments found for $U(1)_{X}$ extensions of
    the $\nu$MSSM where $L_i$ and $H_d$
    are not discriminated ($\nu$MSSM), or where they are ($R_p\nu$MSSM) or
    for the original non-supersymmetric anomaly-free atlas (Non-SUSY). 
    The rightmost column lists the number of potential inequivalent solutions in the SUSY
    case before ACCs
    are applied and where a discrete symmetry distinguishes $L_i$ from $H_d$.
    \label{table:solN}}  
\end{table}

\begin{figure}
  \unitlength = \textwidth
  \begin{center}
    \begin{picture}(0.8,0.5)
    \includegraphics[width=0.8 \textwidth]{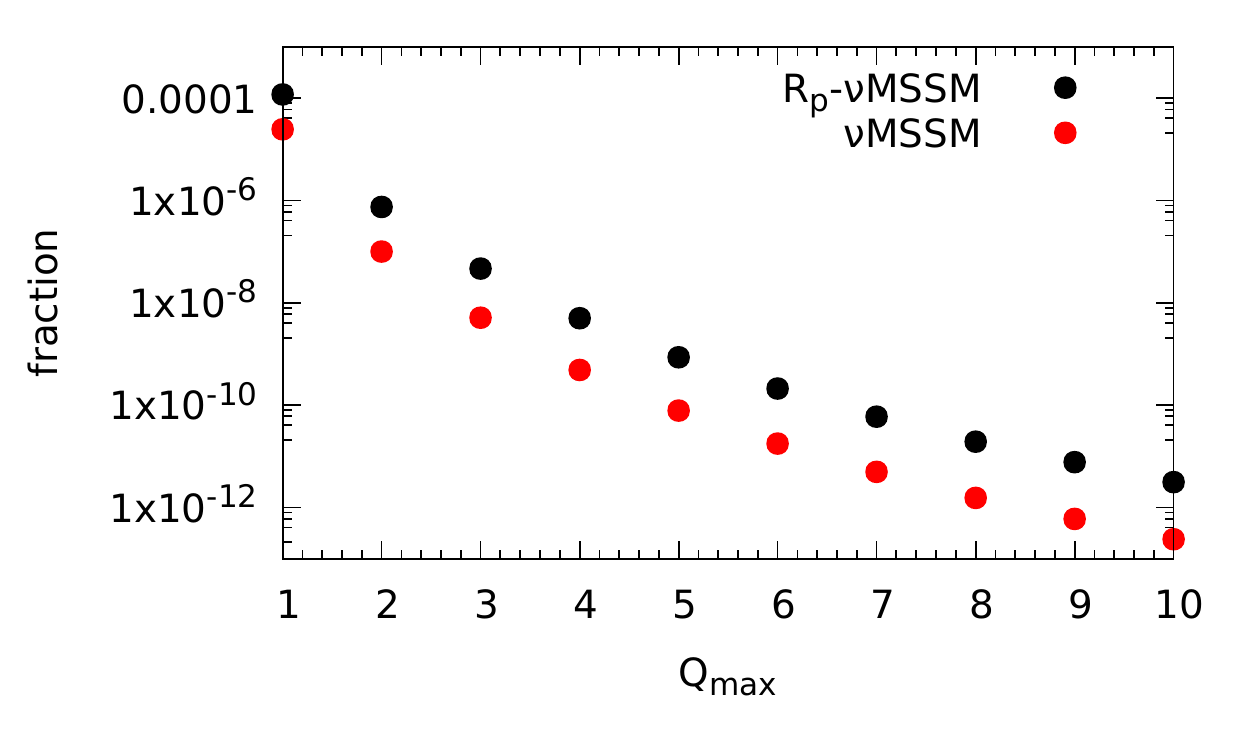}
  \end{picture}
  \caption{Fraction of otherwise possible solutions that are anomaly free as
  a function of $Q_\text{max}$. Specifically, the fraction is equal to the
  number of inequivalent solutions divided by the number of inequivalent possible
  assignments before the local anomaly cancellation
  requirements are imposed. \label{fig:frac}}
  \end{center}
  \end{figure}

We display some basic statistics characterising the number of solutions found in
Table~\ref{table:solN}. The number of inequivalent solutions increases
rapidly as a function of the maximum height searched over, $Q_\text{max}$. In
all, we find over 1.6 
billion solutions for $Q_\text{max}=10$ in the case where a discrete symmetry
does not pick out one of the $X_{L_\alpha}$ charges to be $X_{H_d}$, a far larger
number than the original anomaly-free atlas (which counts under 22 million
inequivalent solutions). As expected, this increases
almost four-fold for the case where one does pick an $X_{L_\alpha}$ to be $H_d$. 
Solutions to the ACCs are scarce; their density decreases with increasing height. For
a height of 10, for example, only approximately 1 in $10^{12}$ possible
inequivalent charge assignments are anomaly free. 
We display this fraction for various different values of $Q_\text{max}$ in
Fig.~\ref{fig:frac} for the case where no symmetry discriminates between $H_d$
and $L_i$ ($\nu$MSSM) and the case where it does ($R_p \nu$MSSM).

\begin{table}
\begin{center}
\setlength\tabcolsep{4pt}
  \begin{tabular}{|c|c|c|c|c|c|c|c|c|c|c|c|c|c|c|c|c|c|c|c|c|} 
  \hline
  \small{Model} & $Q$ & $Q$ & $Q$ & $n$ & $n$ & $n$ & $e$ & $e$ & $e$ & $u$ &
  $u$ & $u$ & $d$ & $d$ & $d$ & $L$ & $L$ & $L$ & $L$ & $\tilde H_{u}$\\
   \hline
    \hline
  $ Y_{3}^\prime$ & $\minus 1$ & $\minus 1$ & $1$ & $0$ & $0$ & $0$ & $\minus 6$ & $6$ & $6$ & $\minus 4$ & $\minus 4$ & $4$ & $\minus 2$ & $2$ & $2$ & $\minus 3$ & $3$ & $3$ & $3$ & $\minus 3$\\
   \hline
  $ B_{3}^\prime$ & $\minus 1$ & $\minus1$ & $1$ & $\minus 3$ & $3$ & $3$ & $\minus 3$ & $3$ & $3$ & $\minus 1$ & $\minus 1$ & $1$ & $\minus 1$ & $\minus1$ & $1$ & $\minus 3$ & $3$ & $3$ & $3$ & $\minus3$\\
   \hline
  Ref. \cite{Demir:2005ti}, Table 3 & $0$ & $0$ & $0$ & $\minus 3$ & $0$ & $0$ & $\minus 1$ & $ 1$ & $3$ & $\minus 1$ & $\minus 1$ & $\minus 1$ & $1$ & $1$ & $1$ & $\minus 2$ & $0$ & $1$ & $2$ & $\minus1$\\
   \hline
     Ref. \cite{Demir:2005ti}, Table 4 & $\minus1$ & $\minus1$ & $\minus1$ & $\minus 9$ & $0$ & $0$ & $\minus 9$ & $0$ & $0$ & $1$ & $1$ & $1$ & $1$ & $1$ & $1$ & $0$ & $0$ & $0$ & $9$ & $0$\\
   \hline
  Ref. \cite{Duan:2018akc} & $\minus 1$ & $0$ & $0$ & $\minus 1$ & $0$ & $4$ & $\minus 1$ & $0$ & $4$ & $\minus 1$ & $0$ & $0$ & $\minus 1$ & $0$ & $0$ & $\minus 1$ & $0$ & $0$ & $4$ & $0$\\
   \hline
     Ref. \cite{sym13020191} & $\minus 1$ & $0$ & $0$ & $1$ & $1$ & $1$ & $1$ & $1$ & $1$ & $\minus 1$ & $0$ & $0$ & $\minus 1$ & $0$ & $0$ & $0$ & $1$ & $1$ & $1$ & $0$\\
   \hline
  \small{SUSY $B \minus  L$} \cite{Barger:2008wn} & $\minus 1$ & $\minus 1$ & $\minus 1$ & $3$ & $3$ & $3$ & $3$ & $3$ & $3$ & $\minus 1$ & $\minus 1$ & $\minus 1$ & $\minus 1$ & $\minus 1$ & $\minus 1$ & $0$ & $3$ & $3$ & $3$ & $0$\\
  \hline
   \small{TFHM}    \cite{Allanach:2018lvl}& $\minus 1$ & $0$ & $0$ & $0$ & $0$ & $0$ & $0$ & $0$ & $6$ & $\minus 4$ & $0$ & $0$ & $0$ & $0$ & $2$ & $0$ & $0$ & $0$ & $3$ & $0$\\
 \hline
\end{tabular}
\end{center}
\caption{Some examples of anomaly-free charge assignments found. Here, we list the ${\mathfrak
    u}(1)_X$ charges of the
  (left-handed or right-handed) chiral fermions of each model. These include the solutions
  $Y_{3}^\prime$, $B_{3}^\prime$ used to derive the analytic solution of \S\ref{sec:analytic}  as
  well as the non-SUSY Third Family Hypercharge (TFHM) solution which we expect to be contained within our list.  Note that all solutions have been rescaled and reordered to satisfy the format of our list as detailed in \S\ref{sec:comp}.}
   \label{table:fields} 
\end{table}
In Table~\ref{table:fields}, we display some solutions that appear in the literature and in our
list. All of the solutions shown were found using the binary search
  algorithm sketched in \S\ref{sec:bin}.
Their presence in the list is a check of some expected and found solutions.
Two solutions ($Y_3^\prime$ and $B_3^\prime$) will be useful for our analytic solution,
which we turn to now.

\section{Analytic Solution \label{sec:analytic}}

In this section, we will first frame our problem in a geometric language that
will facilitate our analytic solution of the ACCs.
We shall then go on to sketch the geometric method by which the solution is
obtained. Then we shall derive the solution in detail algebraically,
eventually providing an explicit parameterisation of the 20 integer charges of
the $\nu$MSSM 
chiral superfields in terms of some integer parameters. We then provide a
right inverse, which, given a solution to the ACCs, returns parameters which will lead to that solution. Such an inverse has the dual
purpose of facilitating checks between the numerical and analytic solutions
and of providing an additional proof that our solution is generic.

\subsection{Geometric framing of the problem}

The ACCs form a set of polynomial equations in the integers - otherwise called
\emph{diophantine equations}. Suppose we take account of only the physical
equivalence defined by scaling (point (\ref{rescale}) in \S\ref{sec:ACC}). It then does
not matter, from a mathematical point of view, whether we use the label $L_4$ or $H_d$ for the relevant chiral superfield; here
we shall choose the latter. We can view the unknown charges as
corresponding to points in the projective space
$\mathrm{P}\mathbb{Q}^{19}$. This is formed by considering 
the charges as living in the rationals $\mathbb{Q}^{20}$, removing the origin
and providing an
  equivalence relation between points in $\mathbb{Q}^{20}$ differing
by rational multiples. The points satisfying the ACCs in $\mathrm{P}\mathbb{Q}^{19}$ are said to form a \emph{projective variety}.

One might
  expect our solution to be parameterised by 14 independent integer-valued parameters (starting
  with 20 and subtracting 6 for the ACCs). However, as we shall see, we shall
  have to add 9 parameters to cover exceptional cases, making the
  total number of integer parameters 23. Our solution
  will then take the form of a map from $\mathbb{Z}^{23}$ to 
  $\mathrm{P}\mathbb{Q}^{19}$ which satisfies the following properties: its
  image is completely within the projective variety, it surjects onto the
  projective variety and its value depends only on the projection onto a
  $\mathbb{Z}^{14}$ subspace of $\mathbb{Z}^{23}$ in all but a few classes of
  exceptional cases. 

Within $\text{P}\mathbb{Q}^{19}$, the ACCs (\ref{cubic}) and (\ref{quadratic}) 
define a cubic and a quadratic hypersurface, respectively (we shall below refer to these as `the
cubic' and `the quadratic', respectively, for brevity).

To solve systems of diophantine equations, number theorists often use a small set of solutions as a tool for finding all solutions. Given our extensive numerical
scan we are in a position to make an attempt in this manner. For a generic set of
equations this is not guaranteed to be possible, however we are lucky in that
for our particular set of ACCs, at least two distinct methods exist. The first
mirrors the method of Ref.~\cite{Allanach:2020zna} which exploits
a special point of P$\mathbb{Q}^{19}$ that is a `double point' of
both the cubic and quadratic. A second new method is presented here. 

\subsection{Sketch of the method}

The linear ACCs are easy to deal with. Their solution defines a projective subspace $\mathrm{PL}$ of $\mathrm{P}\mathbb{Q}^{19}$. It is in PL that we must discuss the quadratic and the cubic.

Given a single solution to the quadratic ACC it is possible to find all solutions
to the quadratic ACC
by constructing all possible lines through this known solution: along
  each line there must be one further solution to the quadratic, since every
  rational quadratic in one dimension has either two or zero rational roots.
In a
similar vein, given a 
single solution to the cubic, with the special property that all first
order partial derivatives vanish at this point, it is possible to find all
solutions to the cubic ACC by constructing lines through this point. Such a
point is called a \emph{double point} of the cubic. 
 
To solve both the quadratic and the cubic simultaneously it is sufficient to
have a \emph{line} on which every point is a solution to the quadratic and
every point is a double point of the cubic (although as noted above, other
methods do  exist). In fact for us there is only one such line (up to
permutations of charges within the $\mathfrak{su}(2)$-doublet,
$\mathfrak{su}(3)$-singlet sector, and other species), which is the one
between the points $Y_3$ and $B_3$ given in Table~\ref{table:solNAna}. These
two points are a reordering of the charges within $Y_3^\prime$ and
$B_3^\prime$, respectively,
from Table~\ref{table:solN}. The first point, $Y_3$,
corresponds to hypercharge except for  the third family, which  has had its
charges sign changed. The second point, $B_3$, corresponds to $B-L$ where the third
family has had its 
charge's {sign changed} and 
the charges  $X_{H_u}$ and $X_{H_d}$ are modified from their usual values of
zero. We will denote the line between them $Y_3B_3$.

To see how $Y_3B_3$ will enable us to find all solutions, let us first define the space $\mathrm{PL}^\prime$, defined to be the subspace of $\mathrm{PL}$ whose points are orthogonal to $Y_3$ and $B_3$ with respect to the standard scalar product on $\mathbb{Q}^{20}$. Every point in $\mathrm{PL}$ lies on a plane $Y_3  B_3 R$ formed by $Y_3B_3$ and a point $R\in \mathrm{PL}^\prime$. Thus,  we can restrict our attention to looking at such planes, and the points within them which satisfy the ACCs.

\begin{figure}[t]
\begin{center}
\begin{tikzpicture}
\node (Y3) [label={[xshift=0cm, yshift=-0.75cm]\footnotesize$Y_3$}]{};
\node (B3) [right=2in of Y3,label={[xshift=0cm, yshift=-0.75cm]\footnotesize$B_3$}]{};
\node (R) [above right=0.7in and 1.75in of Y3,anchor=east,circle, minimum size=1mm,inner sep=0pt,outer sep=0pt, fill=black,draw,label={[xshift=0cm, yshift=-0.5cm]\footnotesize$R$}]{}; 

\node (P) [above right=0.5in and 0.975in of Y3,anchor=east,circle, minimum size=1mm,inner sep=0pt,outer sep=0pt, fill=black,draw,label={[xshift=0cm, yshift=0.05cm]\footnotesize$P_0$}]{}; 

\node (PL) [above left=0.75in and 0.01in of Y3,label={[xshift=0cm, yshift=0cm]\footnotesize$PL$}]{};

\node (A1) [above right=0.7in and 0.5in of Y3]{};
\node (A2) [above right=0.1in and 1.5in of Y3]{};
\node (B1) [above right=0.7in and 1.4in of Y3]{};
\node (B2) [above right=0.1in and 0.2in of Y3]{};

\node (C2) [above right=1in and 0.2in of Y3]{};
\node (C3) [above right=1in and 0.2in of B3]{};
\coordinate (cY3) at (Y3);
\coordinate (cB3) at (B3);
\coordinate (cC2) at (C2);
\coordinate (cC3) at (C3);
\coordinate (cA1) at (A1);
\coordinate (cA2) at (A2);
\coordinate (cB1) at (B1);
\coordinate (cB2) at (B2);

\draw (cY3)--(cB3);
\draw (cC2)--(cY3);
\draw (cB3) -- (cC3);
\draw (cC2) -- (cC3);
\draw (cA1) -- (cA2) node [pos=0.8, above, sloped] (text) {\tiny (in cubic) $L_c$};
\draw (cB1) -- (cB2) node [pos=0.8, above, sloped] (text) {\tiny (in quadratic) $L_q$};;

\node (aY3) [right=0.75in of B3,label={[xshift=0cm, yshift=-0.75cm]\footnotesize$Y_3$}]{};
\node (aB3) [right=2in of aY3,label={[xshift=0cm, yshift=-0.75cm]\footnotesize$B_3$}]{};
\node (aR) [above right=0.7in and 1.75in of aY3,anchor=east,circle, minimum size=1mm,inner sep=0pt,outer sep=0pt, fill=black,draw,label={[xshift=0cm, yshift=-0.5cm]\footnotesize$R$}]{}; 
\node (a) [above left=0.75in and 0.01in of aY3,label={[xshift=0cm, yshift=0cm]\footnotesize{a)}}]{};
\node (aC2) [above right=1in and 0.2in of aY3]{};
\node (aC3) [above right=1in and 0.2in of aB3]{};
\node (aA1) [above right=0.7in and 0.5in of aY3]{};
\node (aA2) [above right=0.1in and 1.5in of aY3]{};
\node (aB) [above right=-0.2in and 0in of aC2]{\tiny (in quadratic)};
\coordinate (caY3) at (aY3);
\coordinate (caB3) at (aB3);
\coordinate (caC2) at (aC2);
\coordinate (caC3) at (aC3);
\coordinate (caA1) at (aA1);
\coordinate (caA2) at (aA2);
\draw (caY3)--(caB3);
\draw (caC2)--(caY3);
\draw (caB3) -- (caC3);
\draw (caC2) -- (caC3);
\draw (caA1) -- (caA2) node [pos=0.8, above, sloped] (text) {\tiny (in cubic) $L_c$};
\node (bY3) [below=1.5in of Y3,label={[xshift=0cm, yshift=-0.75cm]\footnotesize$Y_3$}]{};
\node (bB3) [right=2in of bY3,label={[xshift=0cm, yshift=-0.75cm]\footnotesize$B_3$}]{};
\node (bR) [above right=0.7in and 1.75in of bY3,anchor=east,circle, minimum size=1mm,inner sep=0pt,outer sep=0pt, fill=black,draw,label={[xshift=0cm, yshift=-0.5cm]\footnotesize$R$}]{}; 
\node (b) [above left=0.75in and 0.01in of bY3,label={[xshift=0cm, yshift=0cm]\footnotesize{b)}}]{};
\node (bC2) [above right=1in and 0.2in of bY3]{};
\node (bC3) [above right=1in and 0.2in of bB3]{};
\node (bA) [above right=-0.2in and 0in of bC2]{\tiny (in quadratic \&
  cubic)};
\coordinate (cbY3) at (bY3);
\coordinate (cbB3) at (bB3);
\coordinate (cbC2) at (bC2);
\coordinate (cbC3) at (bC3);
\draw (cbY3)--(cbB3);
\draw (cbC2)--(cbY3);
\draw (cbB3) -- (cbC3);
\draw (cbC2) -- (cbC3);
\node (cY3) [right=0.75in of bB3,label={[xshift=0cm, yshift=-0.75cm]\footnotesize$Y_3$}]{};
\node (cB3) [right=2in of cY3,label={[xshift=0cm, yshift=-0.75cm]\footnotesize$B_3$}]{};
\node (cR) [above right=0.7in and 1.75in of cY3,anchor=east,circle, minimum size=1mm,inner sep=0pt,outer sep=0pt, fill=black,draw,label={[xshift=0cm, yshift=-0.5cm]\footnotesize$R$}]{}; 
\node (c) [above left=0.75in and 0.01in of cY3,label={[xshift=0cm, yshift=0cm]\footnotesize{c)}}]{};
\node (cC2) [above right=1in and 0.2in of cY3]{};
\node (cC3) [above right=1in and 0.2in of cB3]{};
\node (cB1) [above right=0.7in and 1.4in of cY3]{};
\node (cB2) [above right=0.1in and 0.2in of cY3]{};
\coordinate (ccY3) at (cY3);
\coordinate (ccB3) at (cB3);
\coordinate (ccC2) at (cC2);
\coordinate (ccC3) at (cC3);
\coordinate (ccB1) at (cB1);
\coordinate (ccB2) at (cB2);
\draw (ccY3)--(ccB3);
\draw (ccC2)--(ccY3);
\draw (ccB3) -- (ccC3);
\draw (ccC2) -- (ccC3);
\draw (ccB1) -- (ccB2) node [pos=0.6, above, sloped] (text) {\tiny (in
  quadratic \& cubic)  $L_q$};

\end{tikzpicture}
\end{center}
\caption{A schematic of our method. In the generic case (top left), the plane
  contains one further line in the quadratic and one further line in the
  cubic. These intercept at a point $P_0$, which is our new solution to all
  ACCs. The 
  exceptional cases are shown in diagrams (a), (b) and (c). In (a), $L_c$ is a
line of solutions to all ACCs. In (b), the $Y_3B_3R$ is a whole plane of
solutions to all ACCs and in
(c), the line $L_q$ is a line of solutions to all ACCs.}\label{fig:AnalyticSol}
\end{figure}
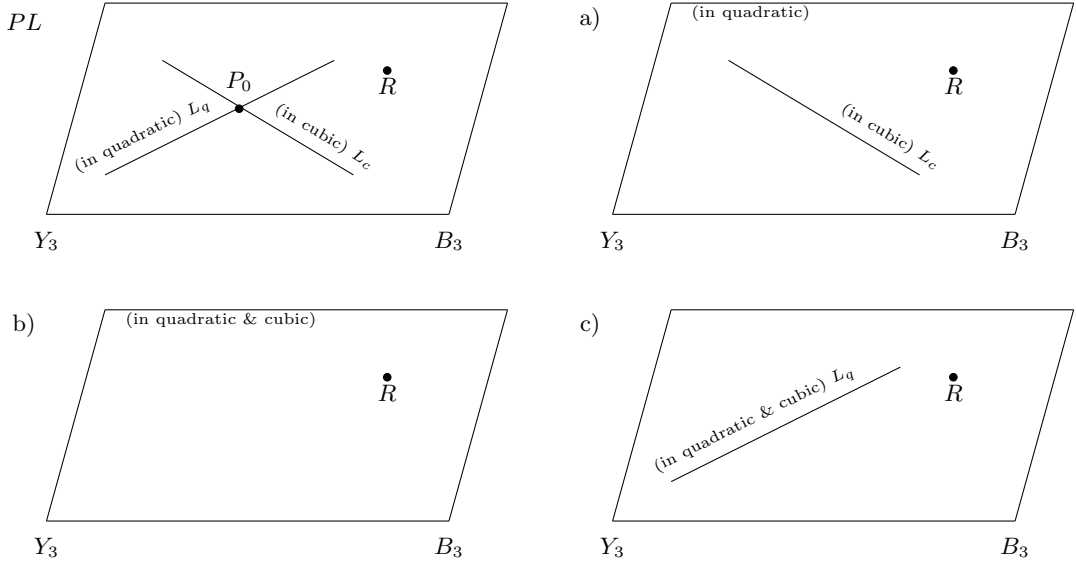

Generically (we will look at the few exceptions shortly), the intersection of
the quadratic with $Y_3  B_3 R$ consists of the union of $Y_3B_3$ and another
line $L_q$, as we will see explicitly in the next subsection. In a similar way, the intersection of the cubic with $Y_3  B_3 R$ consists of the
union of $Y_3B_3$ and another line $L_c$. The intersection of the projective
variety defined by the ACCs, and $Y_3  B_3 R$ then consists of the line
$Y_3B_3$ and a single point which is the intersection of $L_q$ and $L_c$ as
shown in the top left-hand panel of Fig.~\ref{fig:AnalyticSol}. Finding this
point, which is a new solution to the ACCs, is a trivial task, as we 
shall shortly see.
 
Let us now look at the exceptional cases, all of which are illustrated in Fig.~\ref{fig:AnalyticSol}. They correspond to the following situations: (a) the whole plane lies in the quadratic but not the cubic; (b) the whole plane lies in both
the quadratic and the cubic; (c) the whole line $L_q$ lies in the cubic. The asymmetry between
the quadratic and the cubic here is simply a manifestation of the ordering in which we will do our manipulations in the next subsection, and nothing more subtle.

We reiterate that since every point in $\mathrm{PL}$ lies in a plane $Y_3  B_3
R$, by finding all solutions in all such planes (considering either the
generic case or the exceptional cases) we can find every point in the
projective variety. In \S\ref{sec:Analytic_param} we will give an explicit
parameterisation of the solution formed by such considerations. In
\S\ref{sec:inverse} a right-inverse to this parametrisation will be given
explicitly demonstrating its full generality.  
 \begin{table}
 \begin{center}
 \setlength\tabcolsep{4pt}
   \begin{tabular}{|c|c|c|c|c|c|c|c|c|c|c|c|c|c|c|c|c|c|c|c|c|} 
   \hline
   & $Q$ & $Q$ & $Q$ & $n$ & $n$ & $n$ & $e$ & $e$ & $e$ & $u$ & $u$ & $u$ &
   $d$ & $d$ & $d$ & $L$ & $L$ & $L$ & $\tilde H_d$ & $\tilde H_{u}$\\
    \hline
     \hline
   $ Y_{3}$ & $1$ &$1$ &$\minus1$ & $0$ & $0$ & $0$ & $\minus6$ & $\minus6$ & $6$ & $4$ & $4$ & $\minus4$ & $\minus2$ & $\minus2$ & $2$ & $\minus 3$ & $\minus 3$ & $3$ & $\minus3$ & $3$\\
    \hline
   $ B_{3}$ & $1$ & $1$ & $\minus1$ & $\minus3$ & $\minus3$ & $3$ & $\minus3$ & $\minus3$ & $3$ & $1$ & $1$ & $\minus1$ & $1$ & $1$ & $\minus1$ & $\minus3$ & $\minus3$ & $3$ & $\minus3$ & $3$ \\
      \hline
  \hline
 \end{tabular}
 \end{center}
 \caption{A new ordering for the anomaly-free charge assignments $Y_3^\prime$ and $B_3^\prime$ given
   in Table~\ref{table:solN}, adapted for the analytic solution.
   Each row lists the ${\mathfrak
    u}(1)_X$ charges of the
  (left-handed or right-handed) chiral fermions of a model. }
    \label{table:solNAna} 
 \end{table}

\subsection{Derivation of the analytic solution \label{sec:deriv}}
We now give a more detailed description of our solution. To this end we define
\begin{align}
q(X,Z)&:=\sum_{i=1}^3(X_{Q_i}Z_{Q_i}-X_{L_i}Z_{L_i}-2X_{u_i}Z_{u_i}+X_{d_i}Z_{d_i}+X_{e_i}Z_{e_i})+X_{H_u}Z_{H_u}-X_{H_d}Z_{H_d},\nonumber\\
c(W,X,Z)&:=\sum_{i=1}^3(6W_{Q_i}X_{Q_i}Z_{Q_i}+2W_{L_i}X_{L_i}Z_{L_i}-3W_{u_i}X_{u_i}Z_{u_i}-3W_{d_i}X_{d_i}Z_{d_i}-W_{e_i}X_{e_i}Z_{e_i}\nonumber\\&-W_{n_i}X_{n_i}Z_{n_i})+2W_{H_u}X_{H_u}Z_{H_u}+2W_{H_d}X_{H_d}Z_{H_d},
\end{align}
which are respectively derived from the quadratic and cubic ACCs {with 
e.g.\  $X_{Q_1}^3$ replaced with $W_{Q_1}X_{Q_1}Z_{Q_1}$}.  {The maps }$q$ and $c$
are the unique trilinear forms which return, respectively, the quadratic
(\ref{quadratic}) ACC and the
cubic ACC (\ref{cubic}),  when all inputs coincide. 

A point $R\in \mathrm{PL}^\prime$ can be parameterised by the 12 charges $R_{S_1}$ for $S\in \{Q,n,e,u,L,d\}$, $R_{S_2}$ for $S\in \{e,L,d\}$, $R_{d_3}$, $R_{H_u}$ and $R_{H_d}$ as well as an extra two parameters $R_1$ and $R_2$. The remaining charges are given by
\begin{align}
R_{Q_2}&=3(R_1+R_{H_d}+R_{L_1}+R_{L_2}+R_{Q_1})+4R_2+2R_{d_3}+R_{e_1}+R_{e_2},\nonumber\\
R_{Q_3}&=-4(R_1+R_2+R_{Q_1})-2(R_{d_3}+R_{e_1}+R_{e_2})-3(R_{H_d}+R_{L_1}+R_{L_2}),\nonumber\\
R_{n_2}&=4R_1+R_2+3(R_{e_1}+R_{e_2})-R_{n_1}+R_{Q_1},\nonumber\\
R_{n_3}&=-(2R_1+R_2+R_{d_1}+R_{d_2}+R_{d_3}+R_{e_1}+R_{e_2}+R_{Q_1}),\nonumber \\
R_{e_3}&=(R_{d_1}+R_{d_2}+R_{d_3})+3(R_{e_1}+R_{e_2})+4R_1,\nonumber\\
R_{u_2}&=-(6R_1+R_2+R_{d_1}+R_{d_2}+2R_{d_3}+4(R_{e_1}+R_{e_2})+R_{Q_1}+R_{u_1}),\nonumber\\
R_{u_3}&=4R_1+R_2+R_{d_3}+2(R_{e_1}+R_{e_2})+R_{Q_1},\nonumber\\
R_{L_3}&=3(R_1+R_{e_1}+R_{e_2})-(R_{L_1}+R_{L_2})-(R_{H_u}+R_{H_d}).
\end{align}
Substituting the generic point, $\alpha Y_3+\beta B_3+\gamma R$, on the plane $
Y_3  B_3 R$ into the quadratic gives
\begin{align}\label{quad_sub}
\gamma(2\alpha q(Y_3,R)+2\beta q(B_3,R)+\gamma q(R,R))=0.
\end{align}
Putting the exceptional cases to one side for now, this equation generically has two lines of solutions: one specified by $\gamma=0$, namely $Y_3B_3$, and a new line $L_q$, a general point of which is given by
\begin{align} \label{quadlinec}
P_{c_1,c_2,c_3}^{(c)}=&\{c_2q(R,R)-2c_3q(B_3,R)\}Y_3+\{2c_3q(Y_3,R)-c_1q(R,R)\}B_3+ \nonumber
\\&+\{c_1q(B_3,R)-c_2q(Y_3,R)\}R,
\end{align}
where $c_1,c_2,c_3\in \mathbb{Z}$
(over-)parameterise the line.\footnote{Our use of projective space allows us to use, by clearing denominators, $\mathbb{Z}$ for these parameters rather than $\mathbb{Q}$.}

On making the same substitution into the cubic we would get a similar line. However, since we are only interested in the intersection of these two lines, it is sufficient to substitute $P_{c_1,c_2,c_3}^{(c)}$ into the cubic. This yields
\begin{align}
4\{c_1q(B_3,R)-c_2q(Y_3,S)\}^2\big\{&\{2q(B_3,R)c(R,R,R)\-3q(R,R)c(B_3,R,R)\}c_1+\nonumber
\\&\{3q(R,R)c(Y_3,R,R)-2q(Y_3,R)c(R,R,R)\}c_2+\nonumber \\&6\{q(Y_3,R)c(B_3,R,R)-q(B_3,R)c(Y_3,R,R)\}c_3\big\}=0.
\end{align}
Solving for $c_1,c_2,c_3$ generically gives the new solution to the ACCs
\begin{align}
  P_0=\{&3q(R,R)c(B_3,R,R)-2q(B_3,R)c(R,R,R)\}Y_3+\nonumber\\
  \{&2q(Y_3,R)c(R,R,R)-3q(R,R)c(Y_3,R,R)\}B_3+\nonumber \\
  6\{&q(B_3,R)c(Y_3,R,R)-q(Y_3,R)c(B_3,R,R)\}R.
\end{align}

Let us now return to the exceptional cases.
\begin{enumerate}
\item[(a)] The plane lies entirely in the quadratic, but not in the cubic:
  This occurs when $q(Y_3,R)=0$, $q(B_3,R)=0$ and $q(Y_3,R)=0$ but at least
  one of $c(Y_3,R,R)$, $c(B_3,R,R)$ and $c(R,R,R)$ is non-zero. In this case,
  we have a line of solutions (over-)parameterised by $a_1$, $a_2$, and $a_3$
  $\in \mathbb{Z}$
  and given by 
\begin{align}
P^{(a)}_{a_1,a_2,a_3}=\{&a_2 c(R,R,R)-3 a_3c(B_3,R,R)\}Y_3+\{3a_3c(Y_3,R,R)-a_1
c(R,R,R)\}B_3+\nonumber \\3\{&a_1c(B_3,R,R)-a_2 c(Y_3,R,R)\}R.
\end{align}
\item[(b)]  The plane lies entirely within the quadratic and the cubic: This
  occurs when $q(Y_3,R)=0$, $q(B_3,R)$, $q(Y_3,R)=0$, $c(Y_3,R,R)=0$, $c(B_3,R,R)=0$ and
  $c(R,R,R)=0$. In this case, every point on the plane lies in the variety. We
  then parameterise the plane with$b_1,b_2,b_3 \in \mathbb{Z}$:
\begin{align}
P^{(b)}_{b_1,b_2,b_3}=b_1 Y_3+b_2 B_3+b_3 R
\end{align}
\item[(c)]   \sloppy The line in the quadratic and the cubic are the same lines: this occurs (excluding the case where the line is just $\alpha Y_3+\beta B_3$) when \begin{align}
2q(B_3,R)c(R,R,R) &= 3q(R,R)c(B_3,R,R),\nonumber\\ 2q(Y_3,R)c(R,R,R)&=3q(R,R)c(Y_3,R,R), \nonumber\\q(B_3,R)c(Y_3,R,R)&=q(Y_3,R)c(B_3,R,R).\end{align} In this case our solution is the line $P^{(c)}_{c_1,c_2,c_3}$.
\end{enumerate}

It is possible to combine these exceptional cases and the generic case into
one parameterisation of the solution using Kronecker delta functions. This
overall parameterisation is given by
\begin{align}
P=&P_0+\delta_{q(Y_3,R),0}\delta_{q(B_3,R),0}\delta_{q(R,R),0} \{P^{(a)}_{a_1,a_2,a_3}+\delta_{c(Y_3,R,R),0}\delta_{c(B_3,R,R),0}\delta_{c(R,R,R),0} P^{(b)}_{b_1,b_2,b_3}\}+\nonumber\\&\delta_{2q(B_3,R)c(R,R,R),3q(R,R)c(B_3,R,R)}\delta_{2q(Y_3,R)c(R,R,R),3q(R,R)c(Y_3,R,R)}\times\nonumber\\&\delta_{q(B_3,R)c(Y_3,R,R),q(Y_3,R)c(B_3,R,R)} P^{(c)}_{c_1,c_2,c_3}.
\end{align}
This parameterisation is written in terms of the 12 charges and two extra parameters specifying $R$, as well as the parameters $a_1$, $a_2$, $a_3$, $b_1$, $b_2$, $b_3$, $c_1$, $c_2$ and $c_3$, which are needed in the exceptional cases. Taking these parameters to be integers returns an integer-valued solution. 
\subsection{Explicit parameterisation}\label{sec:Analytic_param}
To write the parameterisation more explicitly, we define 
\begin{align}
\Gamma:=&\{3q(R,R)c(B_3,R,R)-2q(B_3,R)c(R,R,R)\}+\nonumber\\&\delta_{q(Y_3,R),0}\delta_{q(B_3,R),0}\delta_{q(R,R),0}\nonumber\\&(a_2c(R,R,R)-3a_3c(B_3,R,R)+\delta_{c(Y_3,R,R),0}\delta_{c(B_3,R,R),0}\delta_{c(R,R,R),0} b_1)\nonumber\\&+\delta_{2q(B_3,R)c(R,R,R),3q(R,R)c(B_3,R,R)}\delta_{2q(Y_3,R)c(R,R,R),3q(R,R)c(Y_3,R,R)}\nonumber\\&\delta_{q(B_3,R)c(Y_3,R,R),q(Y_3,R)c(B_3,R,R)}(c_2 q(R,R)-2c_3q(B_3,R)),\nonumber\\
\Sigma:=&\{2q(Y_3,R)c(R,R,R)-3q(R,R)c(Y_3,R,R)\}+\nonumber\\&\delta_{q(Y_3,R),0}\delta_{q(B_3,R),0}\delta_{q(R,R),0}\times
\nonumber\\& \{3a_3c(Y_3,R,R)-a_1c(R,R,R)+\delta_{c(Y_3,R,R),0}\delta_{c(B_3,R,R),0}\delta_{c(R,R,R),0}b_2\}+\nonumber\\&\delta_{2q(B_3,R)c(R,R,R),3q(R,R)c(B_3,R,R)}\delta_{2q(Y_3,R)c(R,R,R),3q(R,R)c(Y_3,R,R)}\nonumber\\&\delta_{q(B_3,R)c(Y_3,R,R),q(Y_3,R)c(B_3,R,R)}\{2c_3 q(Y_3,R)-c_1q(R,R)\},\nonumber\\
\Lambda:=&6\{q(B_3,R)c(Y_3,R,R)-q(Y_3,R)c(B_3,R,R)\}+\nonumber\\&\delta_{q(Y_3,R),0}\delta_{q(B_3,R),0}\delta_{q(R,R),0}\times\nonumber\\& \{3\{a_1c(B_3,R,R)-a_2c(Y_3,R,R)\}+\delta_{c(Y_3,R,R),0}\delta_{c(B_3,R,R),0}\delta_{c(R,R,R),0}b_3\}+\nonumber\\&\delta_{2q(B_3,R)c(R,R,R),3q(R,R)c(B_3,R,R)}\delta_{2q(Y_3,R)c(R,R,R),3q(R,R)c(Y_3,R,R)}\nonumber\\&\delta_{q(B_3,R)c(Y_3,R,R),q(Y_3,R)c(B_3,R,R)}\{2c_1 q(B_3,R)-2c_2q(Y_3,R)\}.
\end{align}
Then the charges are given explicitly by fourth order polynomials in the
coordinates of $R$:
\begin{alignat}{3}
X_{Q_1}&=\Gamma+\Sigma+\Lambda R_{Q_1},\quad
&&X_{Q_2}=\Gamma+\Sigma+\Lambda R_{Q_2},\quad
&&X_{Q_3}=-\Gamma-\Sigma+\Lambda R_{Q_3},\nonumber\\
X_{n_1}&=-3\Sigma+\Lambda R_{n_1},\quad
&&X_{n_2}=-3\Sigma+\Lambda R_{n_2},\quad
&&X_{n_3}=3\Sigma+\Lambda R_{n_3},\nonumber\\
X_{e_1}&=-6\Gamma-3\Sigma+\Lambda R_{e_1},\quad
&&X_{e_2}=-6\Gamma-3\Sigma+\Lambda R_{e_2},\quad
&&X_{e_3}=6\Gamma+3\Sigma+\Lambda R_{e_3},\nonumber\\
X_{u_1}&=4\Gamma+\Sigma+\Lambda R_{u_1},\quad
&&X_{u_2}=4\Gamma+\Sigma+\Lambda R_{u_2},\quad
&&X_{u_3}=-4\Gamma-\Sigma+\Lambda R_{u_3},\nonumber \\
X_{L_1}&=-3\Gamma-3\Sigma+\Lambda R_{L_1},\quad
&&X_{L_2}=-3\Gamma-3\Sigma+\Lambda R_{L_2},\quad
&&X_{L_3}=3\Gamma+3\Sigma+\Lambda R_{L_3},\nonumber\\
X_{d_1}&=-2\Gamma+\Sigma+\Lambda R_{d_1},\quad
&&X_{d_2}=-2\Gamma+\Sigma+\Lambda R_{d_2},\quad
&&X_{d_3}=2\Gamma-\Sigma+\Lambda R_{d_3},\nonumber\\
X_{H_u}&=3\Gamma+3\Sigma+\Lambda R_{H_u},\quad
&&X_{H_d}=-3\Gamma-3\Sigma+\Lambda R_{H_d}.
\label{parameterisation}
\end{alignat}
\subsection{Right inverse} \label{sec:inverse}
As previously mentioned, this analytic solution has a right inverse, demonstrating
its complete generality. Specifically, let $T$ be a known solution and define
the point $G=108T-(Y_3\cdot T-B_3\cdot T)Y_3-(2B_3\cdot T-Y_3\cdot T)B_3$,
where  `$\cdot$' is the usual scalar product. The point $G$ can be thought of as
$T$ with its components in the line $\alpha Y_3+\beta B_3$ projected out. The parameters $R_{X_j}=G_{X_j}$ (for $X_j$ as above), and
\begin{align}
R_1&=-\sum_{i=1}^3(G_{d_i}+G_{L_i})+G_{e_3}-G_{H_u}-G_{H_d},\nonumber\\
R_2&=\sum_{i=1}^3(G_{d_i}-G_{e_i}+2G_{L_i})-G_{e_3}+2G_{H_u}+2G_{H_d}-G_{Q_1}-G_{n_2},\nonumber\\
a_1&=c(B_3,T,T),\quad
a_2=-c(Y_3,T,T),\nonumber\\
a_3&=-c(B_3,T,T)(Y_3\cdot T-B_3\cdot T)+c(Y_3,T,T)(2B_3\cdot T-Y_3\cdot T).\nonumber\\
b_1&=(Y_3\cdot T-B_3\cdot T)\quad
b_2=(2B_3\cdot T-Y_3\cdot T)\quad
b_3=1\nonumber\\
c_1&=q(B_3,T),\quad
c_2=-q(Y_3,T),\nonumber\\
c_3&=-q(B_3,T)(Y_3\cdot T-B_3\cdot T)+q(Y_3,T)(2B_3\cdot T-Y_3\cdot T),
\label{inverse}
\end{align}
return the point $T$ when substituted into the above analytic solution. In fact, they return $T$ up to a multiplicative constant given by 
\begin{align}
&6\times 108^4(q(B_3,T)c(Y_3,T,T)-q(Y_3,T)c(B_3,T,T))\nonumber\\&+\delta_{q(Y_3,T),0}\delta_{q(B_3,T),0}(3\times 108^3(c(B_3,T,T)^2+c(Y_3,T,T)^2)+108\delta_{c(Y_3,T,T),0}\delta_{c(B_3,T,T),0})\nonumber\\&+2\times 108^2\delta_{q(B_3,T)c(Y_3,T,T),q(Y_3,T)c(B_3,T,T)}(q(B_3,T)^2+q(Y_3,T)^2) \label{multiplicative}
\end{align}
but given that our discussion above has been implicitly in projective space, such multiplicative factors are not relevant.

In the {\tt Zenodo} repository~\cite{zenodo} we provide a {\tt
  Mathematica}\texttrademark\ script 
containing the analytic solution, allowing one to generate solutions
at  will. 
\section{Checks of the solutions \label{sec:checks}}
The material content of \S\ref{sec:comp} is a list of all inequivalent
anomaly-free charge 
assignments 
up to a fixed $Q_\text{max}$. A skeptic could justly ask the question: how
does one know this list is complete without redundancies? The algorithm used
does guarantee it, but one wishes to mitigate potential errors involved
in its computer implementation.
A similar level of scrutiny can be applied to the analytic solution of
\S\ref{sec:analytic}. Although here one might hope the correctness of the solution is
mathematically clear-cut, due diligence requires that we should try to ensure
that no fallacies have been committed. 
Happily, several checks can be carried out to satisfy all but the most fastidious
skeptic. These checks work in three different modes: consistency checks
within the numerical solutions, consistency checks within the analytic
solution alone, and cross-checks between the two. The ability to do
cross-checks between the two is one of several advantages for
providing both. Let us discuss the checks performed for each mode in turn. We
note in passing that all checks were carried out successfully.  

For any computer program, one useful check is to make a second structurally
different program but with the same expected outcome. To this end, we produced a second different program (this one did not use
lexicographic ordering, but instead used an ordering 
similar to that in~Ref.~\cite{Allanach:2018lvl}). The two outputs where then compared
and found to agree. 

The addition of hypercharge to any solution
also leads to a solution, as stated in (\ref{hyper}) of
\S\ref{sec:ACCconditions}. This provides a check of the computer program as 
follows: each solution for a given $Q_{\mathrm{max}}$ had multiples of
hypercharge added or subtracted from it up to three times. If the resulting
 charges had a height less than {or equal to} 10, the binary
search method discussed in \S\ref{sec:comp} was used to confirm that {the solution} was present in
our $Q_\text{max}=10$ list.  

Turning to the analytic solution, the most primitive check is to randomly
choose parameters, generate the corresponding charges and confirm that they
satisfy the ACCs. This check was carried out on $10^5$ randomly generated
solutions. 

The fact that we have a right inverse for our parameterisation means that we
can take a solution, apply the inverse and then the parameterisation to return
another solution. If our analysis is correct this new solution should agree
with the one we started with (up to a scaling). This was carried out on,
again, $10^5$ randomly generated solutions. It was also carried out on all the
{scanned} solutions {in our list} for $Q_{\mathrm{max}}{=10}$, thereby providing the first
cross check between the numerical and analytic solutions. 

The second cross-check between the numerical and analytic solutions was to
generate random solutions using the analytic solution, then to identify those
of height less than or equal to {10} and confirm that these
appear in the numerical solution via the binary search algorithm.

\section{Examples of Filters \label{sec:pheno}}

In this section, we now turn to examples of how our list of solutions to ACCs
 (\ref{lin1})-(\ref{quadratic}) might be 
 filtered in order to identify sets of charge assignments with various
 possible desirable
 phenomenological properties or uses.\footnote{Computer programs implementing
   these filters are 
   available on \texttt{Zenodo}~\cite{zenodo}.}
Note that in what follows, as in \S\ref{sec:analytic}, we will distinguish $H_d$ from
$L_i$, and the number of solutions
satisfying each constraint is therefore to be compared with the second
column of Table~\ref{table:solN}.   

\subsection{The superpotential} \label{sec:superpotential}
In general, interactions between the chiral supermultiplets of the MSSM are
given by the 
superpotential $W = W_{R_{p}} + W_{LV} + W_{BV}$, where 
\begin{equation}
\begin{split}
W_{R_{p}} &= \mu \hat H_{u} \hat H_{d} + (y_{u})_{ij}  \hat U^{c}_{i} \hat Q_{j} \hat H_{u} + (y_{d})_{ij} \hat D^{c}_{i} \hat Q_{j} \hat H_{d} + (y_{e})_{ij}  \hat E^{c}_{i} \hat L_{j} \hat H_{d}, \\
W_{LV} &=  \frac{1}{2} \lambda^{ijk} \hat L_{i} \hat L_{j} \hat E^{c}_{k} +\lambda^{\prime ijk} \hat L_{i} \hat Q_{j} \hat D^{c}_{k}+ \mu^{\prime i} \hat L_{i} \hat H_{u}, \\
W_{BV} &= \frac{1}{2} \lambda^{\prime\prime ijk} \hat U^{c}_{i} \hat D^{c}_{j} \hat D^{c}_{k}. \\
\end{split}\label{eq:superpotential}
\end{equation} 
$\hat{U}^c_i$, $\hat{D}^c_i$, $\hat{Q}_i$, $\hat{L}_i$, $\hat{E}^c_i$, $\hat{H}_{u}$ and $\hat{H}_{d}$ denote the chiral
supermultiplets 
containing of Table~\ref{table:fields}, and we denote flavour
indices by $i,j,k\in \{1,2,3\}$. $\lambda^{ijk}$, $\lambda^{\prime ijk}$,
$\lambda^{\prime\prime ijk}$, $(y_{u,d,e})_{ij}$ are all dimensionless
coupling constants and $\mu, \mu^{\prime i}$ each have mass dimension 1.
Gauge indices have been suppressed. Note that
here we ignore the neutrino chiral supermultiplets $\hat{N}_{i}^c$, postponing their discussion until \S\ref{sec:neutrino}.  Here $W_{R_{p}}$ denotes terms invariant under $R-$parity, whereas $R-$parity is violated in the $L$ and $B$-violating terms $W_{LV}$ and $W_{BV}$ respectively.

\subsubsection{The $\mu$ problem \label{sec:mu}}
The MSSM has a fine tuning problem associated with the $\mu \hat{H}_{u} \hat{H}_{d}$ term.
Given that this term respects supersymmetry and gauge symmetry, there is no
explicitly stated
reason for the scale of $\mu$ to be small.  The gauge group can be extended 
by $U(1)_{X}$ to provide a solution to this so-called $\mu$ problem
\cite{Lee:2007fw}.  This is achieved by charging $\hat{H}_{u}$ and $\hat{H}_{d}$ under
$U(1)_{X}$ such that the $\mu$ term above is forbidden by the $U(1)_{X}$
symmetry.  Instead, the flavon $\theta$ is charged, allowing a term of the form 
(where $h$ is a dimensionless coupling constant)
\begin{equation}
W \supset h \theta \hat{H}_{u} \hat{H}_{d} \rightarrow h \langle \theta \rangle \hat{H}_{u} \hat{H}_{d},
\end{equation}  
such that when the $U(1)_{X}$ symmetry is spontaneously broken, the scalar
component of $\theta$
acquires a vacuum expectation value $\langle \theta \rangle$ at the TeV scale
i.e.\ the $\mu$ term is dynamically generated.\footnote{Further detailed
  model building is required to make sure that $\langle \theta
  \rangle\sim{\mathcal O}(\text{TeV})$, but we shall merely assume here that
  this is possible.} 
The
$\mu\nu$SSM~\cite{Bratchikov:2005vp,Escudero:2008jg,Kpatcha:2019gmq,Lopez-Fogliani:2020gzo}
also solves the $\mu$ problem in precisely this manner.
Any model with such a
dynamically generated $\mu$ term
is often referred to as the
next-to-minimal supersymmetric 
standard model (NMSSM). The NMSSM has received much attention in the
literature~\cite{Ellwanger:2009dp,Maniatis:2009re,Ellis:1988er,King:1995vk}.

Remembering that we shall pick one of the $\hat{N}_i^c$ chiral superfields with a
non-zero charge to be the flavon chiral superfield $\theta^c$, which 
has a non-zero $X$ charge out of necessity,
we search for such solutions in our list of charges by applying the conditions
\begin{equation} \label{eq:muterm}
\exists i\in \{1,2,3\}:\quad X_{H_{u}} + X_{H_{d}}=X_{n_i}\ne 0, 
\end{equation} 
where we take $\theta^{c}$ to be the $\hat{N}_i^c$ superfield which
 satisfies this
 condition.\footnote{The U$\mu\nu$SSM~\cite{Aguilar-Saavedra:2021qbv} uses (\ref{eq:muterm})
 in a certain $U(1)^\prime$ 
 extension of the $\nu$MSSM (involving additional quark fields) to
 solve the $\mu$ problem, also.}
 We find a total of 77 solutions satisfying these constraints with
$Q_\text{max}=1$, constituting $\sim 30 \%$ of the full $Q_\text{max}=1$ list.
This percentage reduces to $20 \%$ when $Q_\text{max}=4$, and $11 \%$ when
$Q_\text{max}=10$, providing in this case a total of $649\,831\,168$ options
for a dynamically generated $\mu$ term. 

\subsubsection{A renormalisable Yukawa sector}
In contrast to the rather weak constraints of (\ref{eq:muterm}),
we may place strong conditions on
the Yukawa sector by requiring that all renormalisable Yukawa couplings of
charged fermions 
are allowed in the superpotential $W_{R_p}$ by being $U(1)_X$ gauge invariant,
i.e.\
they must satisfy the following equations
$\forall i,j\in \{1,2,3\}$:
\begin{equation} \label{eq:yukawa}
 X_{Q_{i}} + X_{H_{u}} - X_{u_{j}} = 0,\qquad
 X_{Q_{i}} + X_{H_{d}} - X_{d_{j}} = 0,\qquad
 X_{L_{i}} + X_{H_{d}} - X_{e_{j}} = 0.
\end{equation}
(\ref{eq:yukawa})
implies family universality for the  species $Q$, $e$, $u$, $L$ and $d$.  For
the non-supersymmetric case, it has been shown that anomaly-free charge
assignments exist which allow all of the renormalisable Yukawa terms
\cite{Allanach:2018vjg}.  One 
can show that in the $\nu$MSSM,
we obtain one solution for each non-supersymmetric solution
of \cite{Allanach:2018vjg}, where we must additionally fix $H_{u}$ and $H_{d}$
to satisfy 
\begin{equation}
3X_{H_{u}} = -3X_{H_{d}} = - 3  \sum_{i=1}^{3} X_{Q_{i}} -\sum_{i=1}^{3}
X_{n_{i}}. \label{addit}
\end{equation} 
(\ref{addit}) means that there cannot be any overlap with the solutions
satisfying (\ref{eq:muterm}), i.e.\ none of these solutions can
simultaneously solve the $\mu$ problem.  By filtering through our list of
charges, we find 2 solutions allowing a fully renormalisable Yukawa sector
with $Q_\text{max}=1$ and 5 with $Q_\text{max}=4$, as shown in Table~\ref{table:yukawaFull}.  The full list of $Q_\text{max}=10$ solutions
comprises 
38 such solutions.

\begin{table}[H]
\begin{center}
\setlength\tabcolsep{4pt}
  \begin{tabular}{|c|c|c|c|c|c|c|c|c|c|c|c|c|c|c|c|c|c|c|c|} 
  \hline
  $Q$ & $Q$ & $Q$ & $n$ & $n$ & $n$ & $e$ & $e$ & $e$ & $u$ & $u$ & $u$ & $d$
  & $d$ & $d$ & $L$ & $L$ & $L$ & $\tilde H_{d}$ & $\tilde H_{u}$\\
   \hline
    \hline
$0$ & $0$ & $0$ & $\minus 1$ & $0$ & $1$ & $0$ & $0$ & $0$ & $0$ & $0$ & $0$ & $0$ & $0$ & $0$ & $0$ & $0$ & $0$ & $0$ & $0$\\
   \hline
$0$ & $0$ &  $0$ & $\minus 1$ & $\minus 1$ &  $\minus 1$ &  $1$  & $1$  & $1$  & $\minus1$ & $\minus 1$ & $\minus 1$ & $1$ & $1$  & $1$ & $0$ & $0$ & $0$ & $1$ & $\minus 1$\\
\hline
$\minus1$  & $\minus1$ & $\minus1$ &$ 3$ &$ 3$& $ 3$ &$ 3$ &$ 3$ &$ 3$ &$\minus1$& $\minus1$ &$\minus1$ &$\minus1$& $\minus1$ &$\minus1$& $ 3$ &$ 3$ &$ 3$ &$ 0$ &$ 0$\\
\hline
$\minus1$ & $\minus1$ & $\minus1$ & $ 2$ & $ 2$ & $ 2$ & $ 4$ & $ 4$ & $ 4$ & $\minus2$ & $\minus2$ & $\minus2$ & $ 0$ & $ 0$ & $ 0$ & $ 3$ & $ 3$ & $ 3$ & $ 1$ & \minus1\\
\hline
$\minus1$ & $\minus1$ & $\minus1$ & $ 4$ & $ 4$ & $ 4$ & $ 2$ & $ 2$ & $ 2$ & $ 0$ & $ 0$ & $ 0$ & $\minus2$ & $\minus2$ & $\minus2$ & $ 3$ & $ 3$ & $ 3$ & $\minus1$ & $ 1$\\
\hline \hline
\end{tabular}
\end{center}
\caption{Anomaly-free charge assignments with $Q_\text{max}=4$ allowing all Yukawa terms at the
  renormalisable level. Each row lists the ${\mathfrak
    u}(1)_X$ charges of the
  (left-handed or right-handed) chiral fermions of a model.  Note that all
  listed solutions satisfy $X_{H_{u}} + X_{H_{d}}= 0$, reducing the ACCs to
  those of the SM after substitution.}
   \label{table:yukawaFull} 
\end{table}

We will now relax the assumption that all Yukawa terms must be present in the
Lagrangian at the renormalisable level.  We will enforce that the top and
bottom quark and the tau lepton tree-level Yukawa terms can be present (since
they are 
closer to order 1 and so more difficult to explain by
non-renormalisable or loop interactions, which imply a suppression below
order 1) 
by applying the constraints
\begin{equation} \label{eq:topyukawa}
\begin{split}
\exists \sigma_1,\sigma_2,\sigma_3,\sigma_4,\sigma_5\in S_3:\quad X_{Q_{\sigma_1(3)}} + X_{H_{u}} - X_{u_{\sigma_2(3)}} &= 0,\\
X_{Q_{\sigma_1(3)}} + X_{H_{d}} - X_{d_{\sigma_3(3)}} &= 0,\\
X_{L_{\sigma_4(3)}} + X_{H_{d}} - X_{e_{\sigma_5(3)}} &=0,
\end{split}
\end{equation} 
where $S_3$ is the group of permutations of $3$ objects. We expect $Q_{\sigma_{1}(3)}$, $u_{\sigma_2(3)}$ and $d_{\sigma_3(3)}$
to be predominantly third generation quarks, and similarly $L_{\sigma_4(3)}$
and $e_{\sigma_5(3)}$ to be predominantly composed of third generation
leptons.  We will further assume that tree-level renormalisable Yukawa terms
are not present for the first and second generation fermions by forbidding all
other terms in the Yukawa matrices.
We can express these constraints
by first defining
\begin{equation}
  P_{ijklmn}:=(X_{Q_i}+X_{H_u}=X_{u_j})\wedge(X_{Q_n}+X_{H_d}=X_{d_k})\wedge
  (X_{L_l}+X_{H_d}=X_{e_m}),
\end{equation}
(where $\wedge$ means logical `and')
and then imposing
\begin{equation} \label{eq:lightyukawa}
({\exists !}\,i,j,k,l,m,n\in \{1,2,3\}: \,P_{ijklmn})
\wedge (\forall i,j,k,l,m,n\in \{1,2,3\}\;P_{ijklmn}\Rightarrow n=i),
\end{equation} 
where, in standard logic notation, ${\exists !}$ means `there exists a unique'.

This choice is made with the fermion mass problem in mind: it allows larger
masses to be generated for the top, 
bottom and tau through the standard Yukawa terms, but forbids them for
the light quarks,  producing a mass hierarchy 
between the light and heavy fermions.  In Ref.~\cite{Demir:2005ti} it was shown that the chiral fermions can obtain their
masses at loop level through the interactions with their superpartners by
including non-holomorphic soft terms in the Lagrangian density.
Alternatively, light fermion masses may be acquired through non-renormalisable
operators after the flavon $\theta$ breaks $U(1)_{X}$.  Either of these
mechanisms require the Lagrangian density to contain terms which will further
constrain the charges.  We shall assume that all first and second generation
fermions acquire their masses through some mechanism such as one of these two,
but leave the 
more model dependent effect of any additional constraints to future
investigations.  

We find that when $Q_\text{max}=1$, the list contains 2 solutions satisfying
the constraints of (\ref{eq:lightyukawa}).  At
$Q_\text{max}=4$ a total of $15\,818$ solutions pass these constraints, and at
$Q_\text{max}=10$ this number grows to $34\,646\,735$.  This makes clear that
by imposing these constraints, not only do we begin to address the fermion
mass problem, but we make way for a larger number of options for
model-building compared to those of a fully allowed renormalisable Yukawa sector.
For example, when $Q_\text{max}=2$ there are 8 solutions which simultaneously
solve the $\mu$ problem and satisfy 
(\ref{eq:lightyukawa}).  This overlap grows to 2\,954 solutions when
$Q_\text{max}=4$ and 4\,088\,200 solutions when $Q_\text{max}=10$.
Furthermore, the constraints
(\ref{eq:lightyukawa}) are inherently flavour non-universal, and thus have the
potential to address the $B$ anomalies.  This overlap will be discussed in more
detail in \S\ref{sec:banomalies}.

\subsubsection{$R-$parity violation}
In contrast to the SM, $L$ and $B$ violating terms are allowed by the field
content and gauge symmetries of the MSSM, as shown in
(\ref{eq:superpotential}).
The simultaneous presence of both $B$ and $L$ violating terms will lead to
proton decay in contravention to experimental bounds
unless one introduces a large degree of fine tuning. Usually, all terms in
  $W_{LV}$ and $W_{BV}$
are forbidden by the imposition of $R-$parity.
In the case that $R-$parity is not imposed though, we may ask that our
  $U(1)_X$ symmetry maintains the stability of
  the proton instead. We can form three broad sets of solution within this
  requirement: where \emph{all} $R-$parity violating terms are banned (this
  will also maintain the stability of the lightest supersymmetric particle,
  which may have the properties to constitute cold dark matter), where
  all terms in $W_{BV}$ are banned but where at least one term in $W_{LV}$ is
  allowed, 
  and those where all terms in $W_{LV}$ are
  banned but at least one in $W_{BV}$ is allowed.
Terms such as those in $W_{LV}$ give a Majorana mass term to left-handed
neutrinos (sometimes through loop diagrams) without the need for right-handed
neutrinos~\cite{Allanach:2007qc}. Terms in $W_{BV}$, on the other hand, can assist in baryogenesis~\cite{Dolgov:2006ay}.

We may ban terms in $W_{BV}$ by imposing $\forall i,j,k \in \{1,2,3 \}$
\begin{equation}
  X_{u_{i}} + X_{d_{j}} + X_{d_{k}} \ne 0
\end{equation}
where $j\ne k$ since the antisymmetry of
$\lambda^{\prime\prime ijk}$ in $j,k$ forbids the $j=k$ terms from appearing
in the superpotential.
Similarly, we may ban all terms in $W_{LV}$ by
imposing the conditions $\forall i,j,k,l,m,n,p \in \{1,2,3 \}$ 
\begin{equation}
X_{L_{i}} + X_{L_{j}} - X_{e_{k}} \ne 0,\qquad
 X_{L_{l}} + X_{Q_{m}} - X_{d_{n}} \ne 0,\qquad
 X_{L_{p}} + X_{H_{u}} \ne 0,
\end{equation}
where $i\ne j$
because $\lambda^{ijk}$ is antisymmetric in $i,j$.  At $Q_{max}=1$ we find 8 solutions which ban all $R-$parity violating terms.  These solutions are listed in Table~\ref{table:BVLV}.  We find a total of 51 solutions which ban $W_{BV}$ while allowing terms in $W_{LV}$.  We find no solutions which ban $W_{LV}$ while allowing terms in $W_{BV}$, i.e.\ the only solutions which ban $W_{LV}$ are those which ban all $R-$parity violation.

\begin{table}[h!]
\begin{center}
\setlength\tabcolsep{4pt}
  \begin{tabular}{| r |c|c|c|c|c|c|c|c|c|c|c|c|c|c|c|c|c|c|c|} 
  \hline
 $Q$ & $Q$ & $Q$ & $n$ & $n$ & $n$ & $e$ & $e$ & $e$ & $u$ & $u$ & $u$ & $d$ &
  $d$ & $d$ & $L$ & $L$ & $L$ & $\tilde H_{d}$ & $\tilde H_{u}$\\
    \hline
$\minus1$ & $\minus1$ & $1$ & $1$ & $1$ & $1$ & $1$ & $1$ & $1$ & $\minus1$ & $\minus1$ & $1$ & $\minus1$ & $\minus1$ & $1$ & $1$ & $1$ & $1$ & $1$ & $1$  \\ 
 \hline

$\minus1$ & $\minus1$ & $1$ & $1$ & $1$ & $1$ & $1$ & $1$ & $1$ & $\minus1$ & $\minus1$ & $1$ & $\minus1$ & $\minus1$ & $1$ & $1$ & $1$ & $1$ & $0$ & $0$  \\ 
 \hline

$\minus1$ & $\minus1$ & $1$ & $1$ & $1$ & $1$ & $1$ & $1$ & $1$ & $\minus1$ & $\minus1$ & $1$ & $\minus1$ & $\minus1$ & $1$ & $1$ & $1$ & $1$ & $1$ & $0$ \\ 
 \hline

$\minus1$ & $\minus1$ & $1$ & $1$ & $1$ & $1$ & $1$ & $1$ & $1$ & $\minus1$ & $\minus1$ & $1$ & $\minus1$ & $\minus1$ & $1$ & $1$ & $1$ & $1$ & $1$ & $\minus1$  \\ 
 \hline

$0$ & $0$ & $0$ & $\minus1$ & $\minus1$ & $\minus1$ & $1$ & $1$ & $1$ & $\minus1$ & $\minus1$ & $\minus1$ & $1$ & $1$ & $1$ & $0$ & $0$ & $0$ & $\minus1$ & $1$  \\ 
 \hline

$0$ & $0$ & $0$ & $\minus1$ & $\minus1$ & $\minus1$ & $1$ & $1$ & $1$ & $\minus1$ & $\minus1$ & $\minus1$ & $1$ & $1$ & $1$ & $0$ & $0$ & $1$ & $1$ & $\minus1$  \\ 
 \hline

$0$ & $0$ & $0$ & $\minus1$ & $\minus1$ & $1$ & $\minus1$ & $1$ & $1$ & $\minus1$ & $\minus1$ & $1$ & $\minus1$ & $1$ & $1$ & $0$ & $0$ & $0$ & $\minus1$ & $1$  \\ 
 \hline

$0$ & $0$ & $0$ & $\minus1$ & $0$ & $0$ & $\minus1$ & $1$ & $1$ & $\minus1$ & $\minus1$ & $1$ & $\minus1$ & $1$ & $1$ & $0$ & $0$ & $0$ & $\minus1$ & $1$  \\ 
   \hline \hline
\end{tabular}
\end{center}
\caption{\label{table:BVLV} At $Q_{max} = 1$, we find 8 anomaly-free charge assignments in our list
  banning all $R-$parity violating terms in the MSSM superpotential. Each row lists the ${\mathfrak
    u}(1)_X$ charges of the
  (left-handed or right-handed) chiral fermions of a model.  }
\end{table}

By increasing the maximum charge $Q_{max}$, we find solutions which ban
$W_{LV}$ while allowing $B$-violation.  At $Q_{max}=10$, 
we find 444\,357\,847 solutions which forbid $W_{LV}$ while allowing terms in $W_{BV}$.  We find 2\,916\,984\,840 solutions which forbid $W_{BV}$ while allowing for terms in $W_{LV}$ at $Q_{max}=10$, and a total of 885\,951\,137 solutions which ban all $R-$parity violating solutions, constituting 14\% of the list of charge assignments.

\subsection{$B$ anomalies} \label{sec:banomalies}
Family-dependent charges in the quark and lepton sectors are well-motivated by
the recent hints at lepton flavour non-universality associated with $b \rightarrow s
\ell^{+} \ell^{-}$ 
transitions~\cite{Altmannshofer:2014cfa,Alonso:2017uky,Bonilla:2017lsq,Bhatia:2017tgo,Ellis:2017nrp,Allanach:2018lvl,Allanach:2019iiy,Greljo:2021xmg,Davighi:2021oel},
also known as `$B$ anomalies'.
Global fits incorporating angular distributions and branching fractions point
towards new physics contributions to the Wilson coefficients $C_{9}$, $C_{10}$
of weak effective theory Hamiltonian operators
$O_{9}$, $O_{10}$, respectively,  where 
\begin{equation} \label{eq:O9O10}
O_{9} = (\bar{s}_{L}' \gamma_{\mu} b'_{L}) (\bar{\mu}' \gamma^{\mu} \mu') \quad O_{10} = (\bar{s}_{L}' \gamma_{\mu} b'_{L}) (\bar{\mu}' \gamma^{\mu} \gamma^{5} \mu').
\end{equation}
Here the primes denote that the
fermionic fields are in the mass eigenbasis. 
A vector-like new physics contribution to $C_{9}$ with $C_{10}=0$, or a
  new physics coupling to left-handed muons through the combination $C_{9} = -C_{10}$, are both
favoured by global fits~\cite{Altmannshofer:2021qrr} in comparison to the
SM.

We will filter through our list in search of solutions potentially capable of
explaining the so-called \textit{$B$ anomalies} via the mediation of
flavour-changing $Z^\prime$ interactions, resulting from the spontaneously
broken $U(1)_X$ symmetry.  We will begin by searching for solutions for which
there exists  $i,j \in \{1,2,3\}$ with $Q_{i}$ and $L_{j}$ charged.  These
will play the role of the left-handed bottom/top quark doublet and muon
respectively, 
contributing to the effective operator $(\bar{b}_{L} \gamma^{\mu} b_{L}) (\mu_{L} \gamma_{\mu}
\mu_{L})+\ldots$ once the heavy $Z^\prime$ is integrated out of the
  effective field theory.  We will assume that a rotation to the mass
eigenbasis will mix the 
down-type quarks such that the necessary $\bar{b}'_{L}\gamma^{\mu} s'_{L}$
coupling is produced.  As well as this, we will require that the left-handed
leptons are not completely flavour universal, i.e. $\exists k \in \{1,2,3\}$
such that $X_{L_{k}} \neq X_{L_{\mu}}$.  This will ensure we can have the necessary
$\mu - e$ flavour non-universality to explain
the $b \rightarrow s \ell^+ \ell^-$ data.

\begin{figure}[h!]
\centering
\includegraphics[width=1\textwidth]{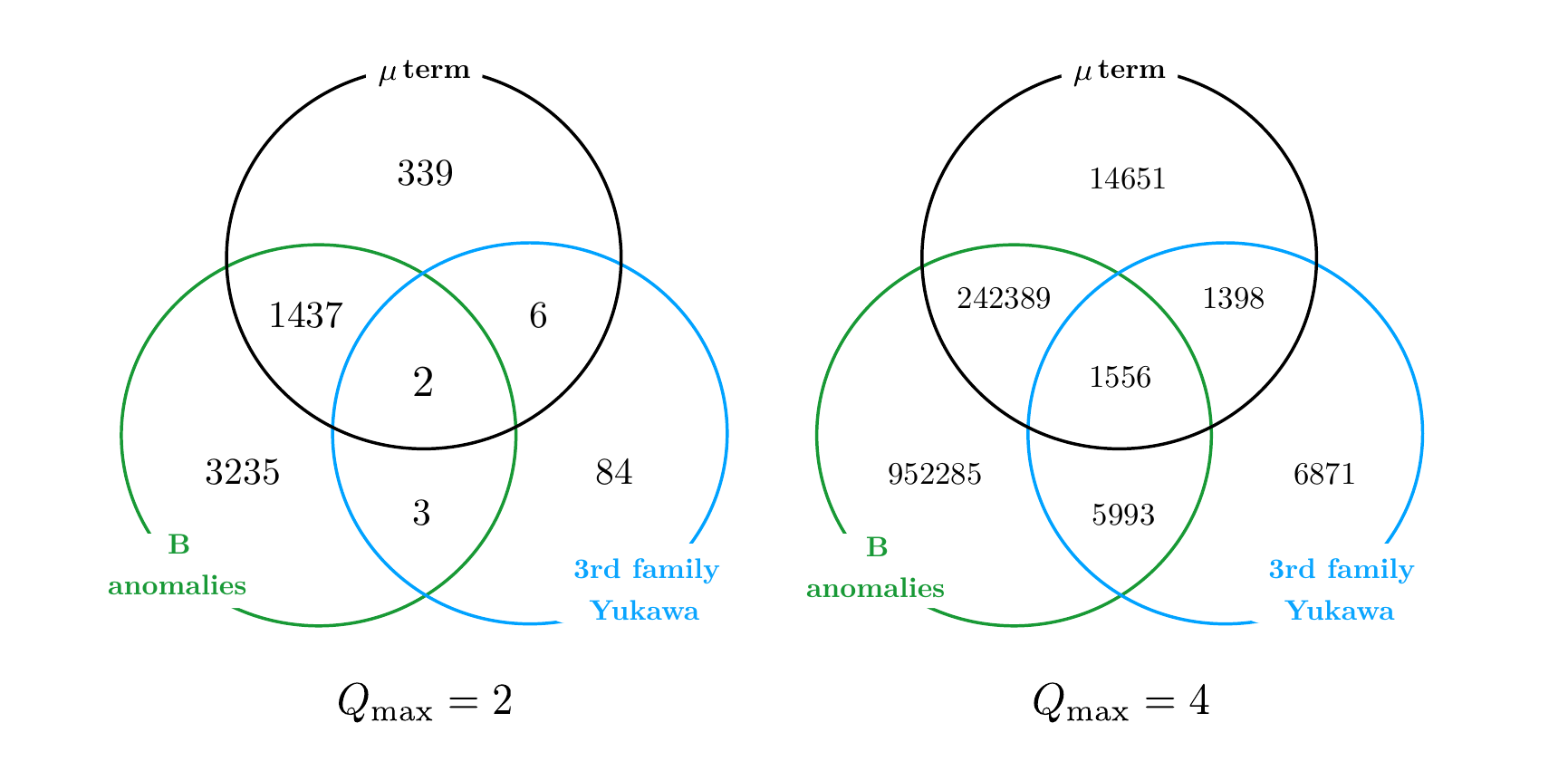}
\caption{We filter through our list to determine the number of solutions capable of solving the $\mu$ problem and the $B$-anomalies, as well as those allowing only 3rd family Yukawa terms.  We find an overlap between these applications, with 2 solutions at $Q_\text{max}=2$ satisfying all constraints and $1\,556$ at $Q_\text{max}=4$.   \label{fig:venn}}
\end{figure}

We find 114 solutions with $Q_\text{max}=1$ satisfying these conditions,
constituting approximately $43 \%$ of the total list.  When $Q_\text{max}=10$
this number grows to $1\,567\,142\,472$, roughly $25 \%$ of the full list of
charge assignments.  Such large numbers indicate that these conditions leave
the charges quite unconstrained, and thus we query the list further for
interesting solutions.  Firstly, there are solutions within this set which can
simultaneously address the $\mu$ problem and allow
only
renormalisable tree-level Yukawa terms
for the top, bottom and tau.  The overlap between each set of constraints is
depicted in Figure~\ref{fig:venn}.  Only 2 solutions can account for all three
conditions when $Q_\text{max}=2$, and are shown in Table~\ref{table:overlapQM1}.  This overlap grows when $Q_\text{max}=4$, with
$1\,556$ solutions solving all three conditions. 

\begin{table}[h!]
\begin{center}
\setlength\tabcolsep{4pt}
  \begin{tabular}{|c|c|c|c|c|c|c|c|c|c|c|c|c|c|c|c|c|c|c|c|} 
  \hline  
$Q$ & $Q$ & $Q$ & $n$ & $n$ & $n$ & $e$ & $e$ & $e$ & $u$ & $u$ & $u$ & $d$ &
  $d$ & $d$ & $L$ & $L$ & $L$ & $\tilde H_{d}$ & $\tilde H_{u}$\\
 \hline
 $\minus 1$ & $0$ & $0$ & $\minus 2$ & $1$ & $2$ & $1$ & $2$ & $2$ & $\minus 2$ & $\minus 1$ & $0$ & $\minus 1$ & $0$ & $2$ & $\minus 1$ & $0$ & $2$ & $1$ & $1$  \\ 
 \hline
$\minus 1$ & $0$ & $0$ & $1$ & $2$ & $2$ & $\minus 2$ & $1$ & $2$ & $\minus 1$ & $0$ & $2$ & $\minus 2$ & $\minus 1$ & $0$ & $\minus 1$ & $0$ & $2$ & $1$ & $1$  \\ 
 \hline
\end{tabular}
\end{center}
\caption{\label{table:overlapQM1}As depicted in Figure~\ref{fig:venn}, at
  $Q_\text{max} = 2$, only 2 anomaly-free charge assignments satisfy the constraints required to
  solve the $\mu$ problem and the $B$-anomalies while allowing 3rd family Yukawa
  terms in the Lagrangian. Each row lists the ${\mathfrak
    u}(1)_X$ charges of the
  (left-handed or right-handed) chiral fermions of a model.}
\end{table}

Secondly, we will filter through the list for solutions that aren't obviously
in danger of violating
experimental constraints.  Following the motivation of Refs.~\cite{Ellis:2017nrp,Allanach:2015gkd}, we search for solutions with uniform
light quark charges so as to avoid constraints on flavour-violation in the
light quark sector.  Additionally, we will search for solutions which feature
zero coupling of the electron to the associated $Z^\prime$ i.e. $\exists$  $i,j \in
\{1,2,3 \}$ such that $X_{L_{i}}=0$ and $X_{e_{j}}=0$.  This is motivated by the
strong experimental constraints originating from $e^+ e^-$ collisions at
  LEP\@.  We find 21 such solutions that also allow only third family Yukawa
terms and address the $\mu$ problem in 
our list with $Q_\text{max}=10$.  A selection of 10 of these solutions are listed
in Table~\ref{table:phenoSols}.  Here, in contrast to other tables, the index
on each fermion denotes the family number (since these are used in the constraints), and we use $\tilde \theta$ to denote
the RH neutrino that plays the role of the (RH) flavino. 

\begin{table}[h!]
\begin{center}
\setlength\tabcolsep{4pt}
  \begin{tabular}{|c|c|c|c|c|c|c|c|c|c|c|c|c|c|c|c|c|c|c|c|c|}   
      \hline
  & $Q_{1}$ & $Q_{2}$ & $Q_{3}$ & $\tilde \theta$ & $n_{1}$ & $n_{2}$ &
      $e_{1}$ & $e_{2}$ & $e_{3}$ & $u_{1}$ & $u_{2}$ & $u_{3}$ & $d_{1}$ &
      $d_{2}$ & $d_{3}$ & $L_{1}$ & $L_{2}$ & $L_{3}$ & $\tilde H_{d}$ &
      $\tilde H_{u}$\\
   \hline
    \hline
  (a) &  $\minus3$ & $\minus3$ & $3$ & $\minus6$ & $0$ & $10$ & $0$ & $9$ & $5$ & $\minus3$ & $\minus3$ & $\minus2$ & $0$ & $0$ & $2$ & $0$ & $9$ & $6$ & $\minus1$ & $\minus5$ \\ 
 \hline
 (b) & $0$ & $0$ & $\minus2$ & $6$ & $\minus3$ & $4$ & $0$ & $3$ & $2$ & $0$ & $0$ & $\minus1$ & $\minus3$ & $\minus3$ & $3$ & $0$ & $3$ & $\minus3$ & $5$ & $1$  \\ 
   \hline
    (c) & $0$ & $0$ & $\minus2$ & $10$ & $\minus1$ & $10$ & $0$ & $0$ & $\minus7$ & $3$ & $3$ & $5$ & $\minus8$ & $\minus8$ & $1$ & $0$ & $6$ & $\minus10$ & $3$ & $7$  \\ 
 \hline
  (d) & $0$ & $0$ & $\minus1$ & $8$ & $\minus9$ & $1$ & $0$ & $0$ & $6$ & $\minus2$ & $\minus2$ & $0$ & $\minus2$ & $\minus2$ & $6$ & $0$ & $\minus4$ & $\minus1$ & $7$ & $1$  \\ 
 \hline
 (e) & $0$ & $0$ & $\minus5$ & $6$ & $8$ & $10$ & $0$ & $\minus1$ & $7$ & $4$ & $4$ & $\minus4$ & $\minus7$ & $\minus7$ & $0$ & $0$ & $7$ & $2$ & $5$ & $1$ \\ 
 \hline
 (f) & $0$ & $0$ & $\minus3$ & $10$ & $\minus9$ & $3$ & $0$ & $8$ & $6$ & $\minus5$ & $\minus5$ & $2$ & $0$ & $0$ & $2$ & $0$ & $\minus2$ & $1$ & $5$ & $5$ \\ 
 \hline
 (g) & $0$ & $0$ & $\minus3$ & $2$ & $0$ & $3$ & $0$ & $7$ & $6$ & $\minus1$ & $\minus1$ & $\minus5$ & $0$ & $0$ & $1$ & $0$ & $5$ & $2$ & $4$ & $\minus2$  \\ 
 \hline
 (h) & $0$ & $0$ & $\minus2$ & $6$ & $\minus6$ & $\minus3$ & $0$ & $8$ & $7$ & $\minus5$ & $\minus5$ & $\minus1$ & $2$ & $2$ & $3$ & $0$ & $\minus2$ & $2$ & $5$ & $1$  \\ 
 \hline
 (i) & $0$ & $0$ & $\minus2$ & $6$ & $\minus6$ & $3$ & $0$ & $4$ & $5$ & $\minus3$ & $\minus3$ & $1$ & $0$ & $0$ & $1$ & $0$ & $\minus2$ & $2$ & $3$ & $3$  \\ 
 \hline
 (j) & $0$ & $0$ & $\minus1$ & $\minus4$ & $\minus6$ & $0$ & $0$ & $7$ & $9$ & $\minus5$ & $\minus5$ & $\minus4$ & $7$ & $7$ & $\minus2$ & $0$ & $\minus3$ & $10$ & $\minus1$ & $\minus3$  \\ 
 \hline
 (k) & $0$ & $0$ & $\minus1$ & $2$ & $2$ & $8$ & $0$ & $\minus5$ & $\minus1$ & $4$ & $4$ & $0$ & $\minus5$ & $\minus5$ & $0$ & $0$ & $3$ & $\minus2$ & $1$ & $1$  \\ 
\hline \hline
\end{tabular}
\end{center}
\caption{At $Q_\text{max}=10$ we find 21 solutions which simultaneously solve
  the $\mu$ problem and $B$ anomalies, allow 3rd family Yukawa terms and are
  well-suited to avoid strong experimental constraints from LEP
  and quark flavour violation between the first two families.  A selection of 10
  of these are shown here. Each row lists the ${\mathfrak
    u}(1)_X$ charges of the
  (left-handed or right-handed) chiral fermions of a model.  }
   \label{table:phenoSols} 
\end{table}

The 10 solutions shown all feature suppressed couplings of the $Z^\prime$ to the light quarks, either because the light RH down-type quarks have zero charge, as in solution (a), or because the light LH quarks have zero charge
as in solutions (b)-(k).  In solutions (a) and (b), the muon has equal RH and LH charge i.e.\  $L_{2} = e_{2}$.
This results in a purely vector-like coupling with $C_{10} = 0$.  Similarly, solutions (c), (d) and (e) 
are particularly interesting in that they all produce negative values of the ratio
$C_{9} / C_{10}$, with (c) and (d) giving exactly $C_{9} = - C_{10}$ and
solution (e) satisfying $C_{9} = -\frac{3}{4} C_{10}$.  In \S\ref{sec:comp} we queried the full list of charge assignments
  in search of known solutions in the literature, listed in Table~\ref{table:fields}.  None of these
  solutions are found in the list of 21 solutions passing our
    constraints: either because they cannot solve the $\mu$ problem and
  address the 3rd family Yukawa terms simultaneously, or because they do not
  satisfy the constraints we impose to  
  facilitate solving the $B$ anomalies.

\subsection{Neutrino masses} \label{sec:neutrino}
Finally, we turn to the neutrinos.  The inclusion of RH neutrinos has allowed us the flexibility to solve the ACCs while simultaneously addressing the phenomenological constraints of \S\ref{sec:superpotential} and \S\ref{sec:banomalies}, as evidenced by the fact that these solutions often have nonzero charges for the RH neutrinos. In particular, this
can be seen from Table~\ref{table:phenoSols} in which all of the solutions
feature nonzero charges for at least one of the RH neutrinos.  It is then
useful to ask what these charge assignments imply for the neutrino masses and mixings.

In order to describe neutrino masses and mixings, we extend the superpotential to include the following terms,
\begin{equation} \label{eq:seesaw}
W = W_{R_{p}} + (y_{\nu})_{ij} N_{i}^{c} L_{j} H_{u} + (M_{\nu_{R}})_{ij} N^{c}_{i} N^{c}_{j},
\end{equation}
where $(y_\nu)_{ij}$ is a 3 by 3 matrix of dimensionless Dirac Yukawa coupling
constants and $(M_{\nu_R})_{ij}$ is a 3 by 3 matrix of Majorana mass terms (of
mass dimension 1) for
the RH neutrinos.
Neutrino masses are then produced through a Type-1 see-saw mechanism.  Many
alternative mechanisms for producing neutrino masses in the MSSM exist in the
literature.  Bilinear $R-$parity violating models extend the superpotential to
include the $L$-violating $\mu'_{i} L_{i} H_{u}$ terms which produce neutrino
masses through mixing with the neutralinos \cite{Hall:1983id,Lee:1984tn}.  In
\cite{Lee:2007qx}, a suppressed Dirac mass term is produced after
$U(1)_{X}$-breaking, through the flavon's vacuum expectation value $\langle
\theta \rangle$. The $\mu \nu SSM$ extends the MSSM to produce neutrino masses
through the inclusion of the trilinear term $\kappa_{ijk} N^{c}_{i} N^{c}_{j}
N^{c}_{k}$ in the
superpotential~\cite{Bratchikov:2005vp,Escudero:2008jg,Kpatcha:2019gmq,Lopez-Fogliani:2020gzo}. 
While an investigation into each of these mechanisms and models is beyond
the scope of this paper, we will filter through our list in search of
solutions which allow all of the terms of (\ref{eq:seesaw}), allowing all
possible neutrino masses
and mixings
via the see-saw mechanism.  These solutions must satisfy the following
constraints 
$\forall i,j\in\{1,2,3\}$
\begin{equation}
X_{L_{i}} + X_{H_{u}} - X_{n_{j}} = 0, \qquad
X_{n_{i}} + X_{n_{j}}  = 0,
\end{equation}
implying $X_{n_i}=0$ and 
$X_{L_i}=-X_{H_u}$.  We find a total of 3 solutions with $Q_\text{max}=1$ in
our list 
satisfying these constraints.  At $Q_\text{max}=4$ a total of $118$ solutions
exist, and at $Q_\text{max}=10$ the list contains 4\ 878 of these solutions.

\subsection{Summary of constraints}
We summarise the phenomenological constraints of this section in Table~\ref{table:phenosummary}.  We emphasise that the filters used throughout this section provide an initial exploration into the constraints we expect will be most commonly needed by model builders.  We expect that the scope of this list is much broader than the phenomenological applications dealt with here, and by making the list of charge assignments publicly available on \texttt{Zenodo}~\cite{zenodo} we encourage model builders to search for charge assignments of more specific interest.

\renewcommand{\arraystretch}{1.1}
\begin{table}[h!]
\begin{center}
\setlength\tabcolsep{3.5pt}
\begin{tabular}{|p{2in}|p{2.9in}|c|}
\cline{2-3}
\multicolumn{1}{c|}{}& Proposition &\# $Q_{\text{max}}=10$\\
\hline
$\mu$ problem & $\exists i\in\{1,2,3\}:$ $X_{H_u}+X_{H_d}=X_{n_i}$ $\wedge$ $X_{n_i}\ne 0$ & $649\,831\,168$ \\ \hline \hline
All renormalisable charged fermion Yukawas & 
$(\forall i,j,k,l,m,n\in \{1,2,3\}\,P_{ijklmn})$ & 38
\\ \hline 
Only 3rd family renormalisable charged fermion Yukawas &$({\exists !}\,i,j,k,l,m,n\in
\{1,2,3\}:\,P_{ijklmn})\wedge$\newline $(\forall i,j,k,l,m,n\in \{1,2,3\}P_{ijklmn}\Rightarrow n=i)$&  34\,646\,735
\\ \hline \hline
$L$-conservation \& $B$-violation & $P_L\wedge\neg P_B$& 444\,357\,847\\
\hline 
$B$-conservation \& $L$-violation &$P_B\wedge \neg P_L$  &  2\,916\,984\,840\\\hline
$L$ \& $B$-conservation & $P_L\wedge P_B$ &885\,951\,137\\ \hline \hline
$B$ anomalies & $\exists i,j,k\in \{1,2,3\}:${}$X_{Q_i}\ne 0$ $\wedge$ $X_{L_j}\ne 0$ $\wedge$ $X_{L_k}\ne X_{L_j}$ & $1\, 567\, 142\, 472$ \\ \hline
	$B$ anomalies, $\mu$ problem, 3rd family Yukawa terms \& experimental constraints & See \S\ref{sec:banomalies} & 21 \\ \hline \hline
See-saw $\nu$ masses & $\forall i,j\in \{1,2,3\} $ $X_{L_i}+X_{H_u}=X_{n_j}$ $\wedge$ $X_{n_i}=-X_{n_j}$&4\,878\\ 
\hline \hline
\end{tabular}
\end{center}
\caption{\label{table:phenosummary} Summary of the phenomenological conditions
  applied in this paper, along with the number of inequivalent $Q_{\text{max}}=10$
  solutions which satisfy them. In the above we have used standard logic
  notation in which $\forall$ reads as `for all', $\wedge$ as `and', $\vee$ as
  `or', $\exists$
  as `there exists', $\exists!$ as `there exists a unique', $\Rightarrow$ as
  `implies', $:$ as `such that', $\neg$ as `not'. For the condition of allowing all
  renormalisable charged fermion
  Yukawa terms, we have used the proposition $P_{ijklm}$ defined as
  $P_{ijklmn}:=(X_{Q_i}+X_{H_u}=X_{u_j}\wedge X_{Q_n}+X_{H_d}=X_{d_k}\wedge
  X_{L_l}+X_{H_d}=X_{e_m})$.  For the $R$-parity related conditions we
    have used the
    propositions $P_L:= \forall i,j,k,l,m,n,p \in \{1,2,3 \} \quad  i= j  \vee(X_{L_{i}} + X_{L_{j}} - X_{e_{k}} \ne 0
\wedge\; X_{L_{l}} + X_{Q_{m}} - X_{d_{n}} \ne 0
\wedge\; X_{L_{p}} + X_{H_{u}} \ne 0)$, and
${P_B}:=\forall i,j,k \in \{1,2,3 \} \quad i=j  \vee X_{u_{i}} + X_{d_{j}} +
X_{d_{k}} \ne 0$.}
\end{table}

\section{Summary}
\label{sec:summary}

Specific models
incorporating the MSSM with an additional $U(1)_X$ gauge group can combine
the phenomenological advantages of supersymmetry with potential uses of the
additional gauge factor and they have received quite some attention in the
literature, particularly for the case where the $U(1)_X$ charges are family dependent.
We have found, for the first time, all charge assignments of the MSSM plus three
SM-singlet chiral superfields which are free of local anomalies (the
SM-singlets can produce neutrino masses as well as spontaneously break the
$U(1)_X$ symmetry).
Chiral superfields in 
real representations can be added to any anomaly-free matter content
and  
result in an anomaly-free solution, since the additional fermionic content
will be in a 
vector-like representation of the gauge group and so its effects cancel in the
anomalies.
The local anomaly 
cancellation conditions described in \S\ref{sec:ACC} constitute a system of six homogeneous coupled 
diophantine equations~(\ref{lin1})-(\ref{eq:ACCs}), the like of which are
notoriously difficult to solve, in general.

Global anomalies are beyond the scope of our work; however, for the case of
$U(1)$ extensions of the usual SM gauge group, there are
none~\cite{Davighi:2019rcd}.
One may question whether a quantum field theory absolutely \emph{has} to be
free from anomalies; 
after all, in an infra-red effective field theory (such as we might expect the
MSSM$\times U(1)_X$ to be) one can in principle add Wess-Zumino terms to the Lagrangian
density in order to cancel them. Such terms can result from decoupling a heavy
state from the effective field theory.
In order to contribute to the anomaly
though, the additional heavy state must be a chiral fermion of non-zero $U(1)_X$
charge. It is then not \emph{a priori} obvious how such a state may acquire a large mass, unless
it is linked to the scale of $U(1)_X$ breaking.\footnote{Integrating
  the top quark out of the SM yields apparent gauge anomalies, but when one
  includes effective operators resulting from integrating it out, gauge symmetry
  is restored~\cite{Preskill:1990fr}. This is precisely a case where the heavy
  mass is linked 
  to the symmetry breaking scale (in this case, of the electroweak
  symmetry).}  
One recent non-supersymmetric $U(1)$ gauge extension of the SM~\cite{Davighi:2021oel} has
achieved this with some additional fermions that under the SM are in vector-like representations,
but which are chiral with respect to $U(1)_X$. However, it is far from obvious
whether this is possible
in 
general model set-ups, particularly when several mixed anomalies do not
cancel. 
From the model builder's point of view
therefore, it is safer to begin with an anomaly-free effective field
theory rather than having to worry about how such anomalies are
cancelled.

We have provided the general analytic solution for the charges via a new
geometric method (a different geometric method was previously employed to solve the
anomaly cancellation conditions for 
non-supersymmetric $U(1)_X$ extensions of the SM~\cite{Allanach:2020zna})
described in \S\ref{sec:analytic}. One
inputs 23 integer parameters for each anomaly-free charge assignment. A {\tt
  Mathematica}\texttrademark\ program 
has been made publicly available which, given the input parameters, produces
one such assignment. The general analytic solution passed
various internal consistency checks. Whilst the general analytic solution can
be difficult for model builders to use, it is useful for (among other things)
providing non-trivial checks of any list of numerical solutions.

Anomaly-free charge assignments are 
scarce: for example, for heights up to 10, as Fig.~\ref{fig:frac} shows, only one out of
some $10^{12}$ (or so) inequivalent assignments is anomaly free.
Despite their scarcity, the different assignments are still legion (we have
identified over 1.6 billion up to a height of 10). The model builder is
therefore faced with an enormous haystack in which to find the proverbial needle. 

An explicit list of all of these 1.6 billion inequivalent charge assignments
up to a 
maximum absolute value of 10 has been
produced via a computer program described in \S\ref{sec:comp} and made
publicly available~\cite{zenodo}. Each
entry in the list comprises 20 integers, the $U(1)_X$ charge assignments of
20 chiral superfields of 
the model. Extensive checks of the list have been made using the analytic
solution as well as those of internal consistency.
With the aid of a computer, such a list is easily and quickly searched and filtered, looking for charge assignments with various desirable
properties. For example, if fewer than three SM-singlets are required for the
model, one can filter the list and find all solutions where one of the
SM-singlet $U(1)_X$ charges is zero. As far as anomalies go, having a zero
charge for the superfield is equivalent to removing it from the model.
We have shown some simple example
filters, looking for
different desirable properties of the charge assignments in \S\ref{sec:pheno}
as a tutorial in their implementation.
We hope that the list will be of use for beyond-the-MSSM builders in terms of
inspiration and phenomenology. 

\section*{Acknowledgements}
We thank
other members of the \emph{Cambridge Pheno Working Group} (and particularly
B~Gripaios) for discussions. 
This work has been partially supported by STFC HEP consolidated grants
ST/P000681/1 and ST/T000694/1. MM acknowledges support from the Schiff Foundation. JTS is partially supported by STFC consolidated grant~ST/S505316/1.

\bibliography{anom}
\bibliographystyle{JHEP}

\end{document}